\documentclass[preprint]{aastex61}

\usepackage{subfigure}
\usepackage{color}

\newcommand{\Lya}{\mbox{Ly$\alpha$}}
\newcommand{\Ha}{\mbox{H$\alpha$}}

\newcommand{\HI}{H\,\textsc{i}}
\newcommand{\cak}{\ion{Ca}{2}~K}
\newcommand{\cah}{\ion{Ca}{2}~H}
\newcommand{\ca}{\ion{Ca}{2}}
\newcommand{\MgII}{\ion{Mg}{2}}
\newcommand{\SiII}{\ion{Si}{2}}
\newcommand{\SiIII}{\ion{Si}{3}}
\newcommand{\SiIV}{\ion{Si}{4}}
\newcommand{\CII}{\ion{C}{2}}
\newcommand{\CIV}{\ion{C}{4}}
\newcommand{\HeII}{\ion{He}{2}}
\newcommand{\HeIIEUV}{\ion{He}{2}~$\lambda$304}
\newcommand{\HeIIFUV}{\ion{He}{2}~$\lambda$1640}

\begin{document}

\title{The MUSCLES Treasury Survey IV: Scaling Relations for Ultraviolet, \cak, and Energetic Particle Fluxes from M dwarfs
        }

\author{Allison Youngblood}
\affiliation{Laboratory for Atmospheric and Space Physics, University of Colorado, 600 UCB, Boulder, CO 80309, USA}
\affiliation{Department of Astrophysical and Planetary Sciences, University of Colorado, 2000 Colorado Ave, Boulder, CO 80305, USA}
\email{allison.youngblood@colorado.edu}

\author{Kevin France}
\affiliation{Laboratory for Atmospheric and Space Physics, University of Colorado, 600 UCB, Boulder, CO 80309, USA}
\affiliation{Department of Astrophysical and Planetary Sciences, University of Colorado, 2000 Colorado Ave, Boulder, CO 80305, USA}
\affiliation{Center for Astrophysics and Space Astronomy, University of Colorado, 389 UCB, Boulder, CO 80309, USA}

\author{R. O. Parke Loyd}
\affiliation{Laboratory for Atmospheric and Space Physics, University of Colorado, 600 UCB, Boulder, CO 80309, USA}
\affiliation{Department of Astrophysical and Planetary Sciences, University of Colorado, 2000 Colorado Ave, Boulder, CO 80305, USA}

\author{Alexander Brown}
\affiliation{Center for Astrophysics and Space Astronomy, University of Colorado, 389 UCB, Boulder, CO 80309, USA}

\author{James P. Mason}
\affiliation{Laboratory for Atmospheric and Space Physics, University of Colorado, 600 UCB, Boulder, CO 80309, USA}

\author{P. Christian Schneider}
\affiliation{European Space Research and Technology Centre (ESA/ESTEC), Keplerlaan 1, 2201 AZ Noordwijk, The Netherlands}

\author{Matt A. Tilley}
\affiliation{Department of Earth \& Space Sciences, University of Washington, Box 351310, Seattle, WA 98195, USA}
\affiliation{NASA Astrobiology Institute -- Virtual Planetary Laboratory Lead Team, USA}
\affiliation{Astrobiology Program, University of Washington, 3910 15th Ave. NE, Box 351580, Seattle, WA 98195, USA}

\author{Zachory K. Berta-Thompson}
\affiliation{Department of Astrophysical and Planetary Sciences, University of Colorado, 2000 Colorado Ave, Boulder, CO 80305, USA}
\affiliation{Center for Astrophysics and Space Astronomy, University of Colorado, 389 UCB, Boulder, CO 80309, USA}

\author{Andrea Buccino}
\altaffiliation{Visiting Astronomer, Complejo Astron\'omico El Leoncito operated under agreement between the Consejo Nacional de Investigaciones Cient\'ificas y T\'ecnicas de la Rep\'ublica Argentina and the National Universities of La Plata, C\'ordoba and San Juan}
\affiliation{Dpto. de F\'isica, Facultad de Ciencias Exactas y Naturales (FCEN), Universidad de Buenos Aires (UBA), Buenos Aires, Argentina}
\affiliation{Instituto de Astronom\'ia y F\'isica del Espacio (IAFE, CONICET-UBA), Casilla de Correo 67, 1428, Buenos Aires, Argentina}

\author{Cynthia S. Froning}
\affiliation{Department of Astronomy/McDonald Observatory,  C1400, University of Texas at Austin, Austin, TX 78712, USA}

\author{Suzanne L. Hawley}
\affiliation{Astronomy Department, Box 351580, University of Washington, Seattle, WA 98195, USA}

\author{Jeffrey Linsky}
\affiliation{JILA, University of Colorado and NIST, 440 UCB, Boulder, CO 80309, USA}

\author{Pablo J. D. Mauas}
\affiliation{Dpto. de F\'isica, Facultad de Ciencias Exactas y Naturales (FCEN), Universidad de Buenos Aires (UBA), Buenos Aires, Argentina}
\affiliation{Instituto de Astronom\'ia y F\'isica del Espacio (IAFE, CONICET-UBA), Casilla de Correo 67, 1428, Buenos Aires, Argentina}

\author{Seth Redfield}
\affiliation{Astronomy Department and Van Vleck Observatory, Wesleyan University, Middletown, CT 06459, USA}

\author{Adam Kowalski}
\affiliation{National Solar Observatory, University of Colorado Boulder, 3665 Discovery Drive, Boulder, CO 80303, USA}
\affiliation{Department of Astrophysical and Planetary Sciences, University of Colorado, 2000 Colorado Ave, Boulder, CO 80305, USA}

\author{Yamila Miguel}
\affiliation{Observatoire de la Cote d'Azur, Boulevard de l'Observatoire, CS 34229 06304 NICE Cedex 4, France}

\author{Elisabeth R. Newton}
\affiliation{Massachusetts Institute of Technology, 77 Massachusetts Ave, Cambridge, MA 02138}

\author{Sarah Rugheimer}
\affiliation{School of Earth and Environmental Sciences, University of St. Andrews, Irvine Building, North Street, St. Andrews, KY16 9AL, UK}

\author{Ant\'igona Segura}
\affiliation{NASA Astrobiology Institute -- Virtual Planetary Laboratory Lead Team, USA}
\affiliation{Instituto de Ciencias Nucleares, Universidad Nacional Aut\'onoma de M\'exico}

\author{Aki Roberge}
\affiliation{NASA Goddard Spaceflight Center, Greenbelt, MD 20771, USA}

\author{Mariela Vieytes}
\affiliation{Dpto. de F\'isica, Facultad de Ciencias Exactas y Naturales (FCEN), Universidad de Buenos Aires (UBA), Buenos Aires, Argentina}

\received{February 27, 2017}
\revised{May 5, 2017}
\accepted{May 11, 2017}
\submitjournal{ApJ}

\begin{abstract}

Characterizing the UV spectral energy distribution (SED) of an exoplanet host star is critically important for assessing its planet's potential habitability, particularly for M dwarfs as they are prime targets for current and near-term exoplanet characterization efforts and atmospheric models predict that their UV radiation can produce photochemistry on habitable zone planets different than on Earth. To derive ground-based proxies for UV emission for use when \textit{Hubble Space Telescope} observations are unavailable, we have assembled a sample of fifteen early-to-mid M dwarfs observed by \textit{Hubble}, and compared their non-simultaneous UV and optical spectra. We find that the equivalent width of the chromospheric \cak~line at 3933 \AA, when corrected for spectral type, can be used to estimate the stellar surface flux in ultraviolet emission lines, including \HI~\Lya. In addition, we address another potential driver of habitability: energetic particle fluxes associated with flares. We present a new technique for estimating soft X-ray and $\textgreater$10 MeV proton flux during far-UV emission line flares (\SiIV~and \HeII) by assuming solar-like energy partitions. We analyze several flares from the M4 dwarf GJ 876 observed with \textit{Hubble} and \textit{Chandra} as part of the MUSCLES Treasury Survey and find that habitable zone planets orbiting GJ 876 are impacted by large Carrington-like flares with peak soft X-ray fluxes $\geq$~10$^{-3}$ W m$^{-2}$ and possible proton fluxes $\sim$10$^{2}$--10$^{3}$ pfu, approximately four orders of magnitude more frequently than modern-day Earth.

\end{abstract}
\keywords{stars: low-mass --- stars: chromospheres --- Sun: flares}

\section{Introduction} \label{sec:Introduction}

\subsection{UV and \cak~emission from M dwarfs}

Recent ultraviolet (UV) studies have shown that even optically-inactive M dwarfs (i.e., those displaying \Ha~spectra in absorption only) display evidence of chromospheric, transition region, and coronal activity \citep{France2013,Shkolnik2014,France2016} that may significantly affect heating and chemistry in the atmospheres of orbiting exoplanets (e.g., \citealt{Segura2003,Segura2005,Miguel2015,Rugheimer2015b,Arney2017}). The Measurements of the Ultraviolet Spectral Characteristics of Low-mass Exoplanetary Systems (MUSCLES) Treasury Survey \citep{France2016} observed 7 nearby ($d$ $\textless$~15 pc), optically-inactive M dwarfs with known exoplanets using the \textit{Hubble Space Telescope} (\textit{HST}), \textit{Chandra}, and \textit{XMM-Newton}. All display quiescent UV emission lines and soft X-rays (SXRs) that trace hot ($T\,\textgreater\,$30,000 K) plasma in the upper stellar atmosphere \citep{Loyd2016}. UV flares were observed from each M dwarf except GJ 1214, the faintest target\footnote{The smaller flares from the MUSCLES M dwarfs show factor of $\textless$10 flux increases, which are below the S/N threshold of GJ 1214's light curves (\citealt{Loyd2017inprep}, in preparation).}.

An M dwarf's far-UV (912--1700 \AA) to near-UV (1700--3200 \AA) spectrum is primarily composed of emission lines that form in the stellar chromosphere and transition region, with a few lines originating in the corona. There is comparitively little continuum emission due to the cool stellar photosphere ($T_{\rm eff}$ $\textless$~4000 K). The \HI~\Lya~emission line (1215.67 \AA) is prominent, comprising 27\%--72\% of the total 1150--3100 \AA~flux (excluding 1210--1222 \AA) for the 7 MUSCLES M dwarfs \citep{France2016,Youngblood2016}. The extreme-UV (100--912 \AA) stellar spectrum is currently not observable in its entirety. No current astronomical observatory exists to observe the 170--912 \AA~spectral range, and the $\sim$400--912 \AA~range is heavily attenuated by neutral hydrogen for all stars, except the Sun. Thus, the extreme-UV must be estimated from other proxies such as \Lya~\citep{Linsky2014} or SXRs~\citep{Sanz-Forcada2011,Chadney2015}. For the M3 dwarf GJ 436, these two methods produce integrated extreme-UV fluxes that agree within 30\% \citep{Ehrenreich2015,Youngblood2016}.

\begin{deluxetable}{cccccccc}
\tabletypesize{\scriptsize}
\tablecolumns{8}
\tablewidth{0pt}
\tablecaption{ The \textit{GOES} classification scheme of solar flares$^a$ \label{table:GOES classification} }
\tablehead{\colhead{Class} & 
                  \multicolumn{2}{c}{Peak flux at 1 AU} & 
                  \multicolumn{2}{c}{SXR Luminosity} & 
                  \colhead{Solar occurrence} &
                  \colhead{Probability of} &
                  \colhead{Expected peak $\textgreater$10 MeV}
                  \\
                  \colhead{} & 
                  \colhead{(W m$^{-2}$)} & 
                  \colhead{(erg cm$^{-2}$ s$^{-1}$)} & 
                  \colhead{(W)} &
                  \colhead{(erg s$^{-1}$)} &
                  \colhead{rate (hr$^{-1}$)$^b$} & 
                  \colhead{CME$^c$} &
                  \colhead{proton flux (pfu)$^d$ at 1 AU}
                  }
\startdata
X100 & 10$^{-2}$ & 10$^{1}$ & 2.8$\times$10$^{21}$ & 2.8$\times$10$^{28}$ & $\textless$2$\times$10$^{-6}$ $^e$ & -- & -- \\
X10 & 10$^{-3}$ & 10$^0$ & 2.8$\times$10$^{20}$ & 2.8$\times$10$^{27}$ & 2$\times$10$^{-5}$ $^f$ & $\sim$100\% & 3300\\
X & 10$^{-4}$ & 10$^{-1}$ & 2.8$\times$10$^{19}$ & 2.8$\times$10$^{26}$ & 0.002 & 80--100\% & 90\\
M & 10$^{-5}$ & 10$^{-2}$ & 2.8$\times$10$^{18}$ & 2.8$\times$10$^{25}$ &  0.02 & 40--80\% & 2\\
C & 10$^{-6}$ & 10$^{-3}$ & 2.8$\times$10$^{17}$ & 2.8$\times$10$^{24}$ &  0.15 & $\textless$40\% & $\textless$1$^f$\\
B & 10$^{-7}$ & 10$^{-4}$ & 2.8$\times$10$^{16}$ & 2.8$\times$10$^{23}$ &  $\textgreater$0.15$^g$ & -- & -- \\
\enddata
\tablenotetext{a}{Classifications are based on the uncorrected \textit{GOES} peak flare flux in the long (1--8 \AA) band at 1 AU. For example, a C3.5-class flare has a peak flux of 3.5$\times$10$^{-6}$ W m$^{-2}$ at 1 AU and an X20-class flare has a peak flux of 2$\times$10$^{-3}$ W m$^{-2}$.}
\tablenotetext{b}{\cite{Veronig2002}.}
\tablenotetext{c}{\cite{Yashiro2006}.}
\tablenotetext{d}{\cite{Cliver2012a}; 1 pfu = 1 proton cm$^{-2}$ s$^{-1}$ sr$^{-1}$.}
\tablenotetext{e}{None have been observed during the \textit{GOES} era (1976--present). }
\tablenotetext{f}{Between 1991 August 25 and 2017 February 6, only 5 flares $\geq$X10-class were observed.}
\tablenotetext{g}{Below the sensitivity limits of \textit{GOES} detectors.}
\end{deluxetable}

Knowledge of an exoplanet's radiation environment is critical for modeling and interpreting its atmosphere and volatile inventory. Specifically, the spectral energy distributions (SEDs) of the host stars are essential, because molecular and atomic cross sections are strongly wavelength dependent. High-energy stellar flux heats upper planetary atmospheres and initiates photochemistry (e.g., \citealt{Lammer2007,Miguel2015,Rugheimer2015b,Arney2017}). UV-driven photochemistry can produce and destroy potential biosignatures (O$_2$, O$_3$, and CH$_4$) and habitability indicators (H$_2$O and CO$_2$) in exoplanet atmospheres \citep{Hu2012,Domagal-Goldman2014,Tian2014,Wordsworth2014,Gao2015,Harman2015,Luger2015b}. In particular, the ratio of far- to near-UV flux determines which photochemical reactions will dominate and thus the resultant planetary atmosphere. Compared to the Sun, M dwarfs have a far-UV to near-UV flux ratio ($F$(1150--1700 \AA)/$F$(1700--3200 \AA)) 100--1000 times larger, and thus most of the known O$_2$ false positive mechanisms predominately impact planets orbiting M dwarfs. Accurately measuring the intrinsic \Lya~flux is critical, because \Lya~comprises the majority of the far-UV flux. For an atmosphere with 0.02 bars of CO$_2$ similar to that simulated by \cite{Segura2007} with GJ 876's SED \citep{France2012,Domagal-Goldman2014}, a 20\% increase (decrease) in the \Lya~flux increases (decreases) the abiotic O$_2$ and O$_3$ column depths by nearly 30\%.

M dwarfs are also prime targets for current and upcoming exoplanet searches and characterization efforts (see \citealt{Scalo2007} and \citealt{Shields2016} for comprehensive overviews) due to their ubiquity in the solar neighborhood \citep{Henry2006}, high occurrence rates of small exoplanets \citep{Dressing2015}, and the larger transit and radial velocity signals their planets provide. The important UV region of an M dwarf's SED cannot yet be predicted by models, although semi-empirical modeling efforts are underway for individual stars (see \citealt{Fontenla2016} and references therein). Thus, direct UV observations of individual stars are currently necessary to model and interpret planetary atmospheric observations, and the characterization of the high-energy SEDs of M dwarfs across a broad range of masses and ages is a community priority (e.g., \citealt{Shkolnik2014b,France2016,Guinan2016}).

Accurate UV spectral flux data is vital in understanding the salient physics and chemistry in an exoplanet atmosphere as well as to break potential degenerate solutions in retrieval models. After \textit{HST} and before future UV observatories begin operations, there will likely be a decade-long gap in UV observing capabilities. This gap may coincide with the majority of \textit{TESS}'s and \textit{CHEOPS}'s habitable zone (HZ) planet detections and subsequent study with \textit{JWST}, so establishing a method of estimating UV spectral flux from ground-based proxies is imperative.

In the optical, the \ca~resonance lines at 3933 \AA~(\cak) and 3968 \AA~(\cah), the \ion{Na}{1} D resonance lines at 5890 and 5896 \AA, \Ha~at 6563 \AA, and the \ca~infrared triplet (IRT) at 8498, 8542, and 8662 \AA~are known to be good indicators of stellar chromospheric activity (e.g., \citealt{Walkowicz2009,GomesdaSilva2011,Lorenzo-Oliveira2016}, and references therein). However, \Ha~and \ion{Na}{1} D are only good indicators of stellar activity for very active stars \citep{Cincunegui2007,Diaz2007,Walkowicz2009} and will not be considered here, because many of our target stars are optically-inactive. The contrast between the \ca~H \& K absorption and emission cores is larger than for the \ca~IRT, so we have limited the scope of this paper to include only one of the \ca~resonance lines. We focus on the \cak~line at 3933 \AA~and ignore the \cah~line at 3968 \AA, because H$\epsilon$ is 1.6 \AA~redward and contaminates \cah~at low spectral resolution. 

The \cak~line profile is a superposition of broad ($\textgreater$1 \AA) absorption and narrow ($\textless$0.5 \AA) emission. In the cooler upper photosphere, Ca$^+$ absorbs against the photospheric continuum, and in the hot upper chromosphere, Ca$^+$ emits. In the cool photospheres and lower chromospheres of M dwarfs, Ca is mostly neutral, so the Ca$^+$ absorption is much narrower than for solar type stars \citep{Fontenla2016}. The Ca$^+$ emission traces stellar activity and has historically been measured for main-sequence stars using the Mt. Wilson S-index \citep{Wilson1978,Vaughan1978} and $R'_{\rm HK}$ \citep{Noyes1984}. The latter corrects the S-index for a spectral type dependence. Other methods involve isolating the emission core from the absorption by fitting and subtracting a non-LTE radiative equilibrium model to the observed spectrum \citep{Walkowicz2009,Lorenzo-Oliveira2016}. 

For the Sun, \cak~is known to correlate well with far-UV emission lines over the 11-year solar cycle and to trace short timescale variation (e.g., flares; \citealt{Tlatov2015}). \cak~originates from solar features like plages and chromospheric network (\citealt{Domingo2009} and references therein) and thus correlates well with the line-of-sight unsigned magnetic flux density (e.g., \citealt{Schrijver1989}). \ca~has also been observed to correlate with \HI~\Lya~for M dwarfs \citep{Linsky2013} despite significant differences between the atmospheric structures of G and M dwarfs (e.g., \citealt{Mauas1997,Fontenla2016}) and potentially different dynamo mechanisms (e.g. \citealt{Chabrier2006,Dobler2006,Browning2008,Yadav2015}). Other known M dwarf scaling relations include SXR--\cak, H$\alpha$--\CIV, H$\alpha$--SXR, H$\alpha$--\MgII, \cak--\MgII, and correlations between \cak~and various Balmer lines \citep{Butler1988,Hawley1991,Hawley2003a,Walkowicz2009}. We improve on these past UV--\ca~scaling relations by increasing the size and diversity (e.g., spectral type and magnetic activity) of the M dwarf sample and expanding to more UV emission lines.

\subsection{Flares and energetic particles}

Habitability studies of M dwarf exoplanets are beginning to include estimates of stellar energetic particles (SEPs; \citealt{Segura2010,Ribas2016}). SEPs can be accelerated by impulsive flares where particles pass along open magnetic field lines into interplanetary space and by shock fronts associated with coronal mass ejections (CMEs; \citealt{Harra2016}).  SEP enhancements and steady-state stellar winds can contribute significantly to planetary atmospheric loss processes by compressing an exoplanet's magnetosphere (e.g., \citealt{Cohen2014,Tilley2016}) and stripping atmospheric particles. SEPs, like high-energy photons, can also catalyze atmospheric chemistry. \cite{Segura2010} found that without the inclusion of particles, large UV flares do not have a long-lasting impact on a HZ planet's O$_3$ column density. However, including SEPs, the expected NO$_{x}$ production will deplete a planet's O$_3$ by $\sim$95\%, requiring centuries for the O$_3$ column density to re-equilibrate after a flaring event ends. This can allow harmful \citep{Voet1963,Matsunaga1991,Tevini1993,Kerwin2007} or bio-catalyzing \citep{Senanayake2006,Barks2010,Ritson2012,Patel2015,Airapetian2016} UV radiation to penetrate to the surface.

Direct measurements of an M dwarf's energetic particle output are not currently possible, but signatures of particle acceleration in the UV and radio should be detectable in principle. Coronal dimming, when extreme-UV emission lines dim after part of the corona has been evacuated from a CME \citep{Mason2014}, was not observed by the \textit{Extreme UltraViolet Explorer (EUVE)}, likely due to insufficient sensitivity. Type II radio bursts that trace shocks associated with CMEs \citep{Winter2015} are being searched for but have not yet been detected on other stars \citep{Crosley2016}, but other possible kinematic signatures of CMEs in observed M dwarf flares have been detected (e.g., \citealt{Houdebine1990,Cully1994,Fuhrmeister2004}). Type III radio bursts are caused by the acceleration of suprathermal electrons from solar active regions and have been detected on the M3 dwarf AD Leo \citep{Osten2006}. Probing the astrosphere (analogous to the heliosphere) via high-resolution \Lya~measurements allows for time-averaged measurements of the stellar mass loss rate \citep{Wood2005a}, which includes the accumulation of impulsive events (CMEs) and the quiescent stellar wind. However, it is uncertain if the kilo-Gauss surface magnetic fields of M dwarfs would allow the acceleration of particles into the astrosphere \citep{Osten2015,Drake2016}. \cite{Vidotto2016} find that rapidly-rotating stars may have strong toroidal magnetic fields that could prevent stellar mass loss, and \cite{Wood2014} find that the scaling relation between SXR flux and mass-loss rate breaks down for stars with SXR surface flux $\textgreater$10$^6$ erg cm$^{-2}$ s$^{-1}$, where a fundamental change in the magnetic field topology may occur. On the Sun in October 2014, the large active region 2192 emitted many large flares, but no CMEs. Strong overlying magnetic fields likely confined the eruption \citep{Thalmann2015a,Sun2015}. As more exoplanetary atmosphere models consider the important influence of stochastic flares \citep{Lammer2007,Segura2010,Airapetian2016,Venot2016}, improving constraints on the fluxes and energies of associated particles is necessary.

To estimate SEPs for any star, we must currently rely on solar relations between SEPs and flare emission \citep{Belov2007,Cliver2012a,Osten2015}. However, traditional flare tracers in the SXR, UV, and U-band (3660 \AA) originate from thermally-heated plasma and probably do not trace particle acceleration processes; CME particles appear to be drawn non-thermally from cooler plasma in the ambient corona \citep{Hovestadt1981,Sciambi1977}. Yet, correlations between flare tracers and SEPs have been detected, likely due to Big Flare Syndrome \citep{Kahler1982}. Big Flare Syndrome explains positive correlations between flare observables that do not share an identified physical process; for a larger total energy release from a flare, the magnitude of all flare energy manifestations will statistically be larger as well. Big Flare Syndrome occurs because a multitude of energy transport mechanisms between layers of the solar atmosphere makes energy transport efficient. These mechanisms include thermal conduction, radiation, bulk convection, and electron condensation and evaporation. For example, \cite{Belov2007} found a correlation between the Sun's 1--8 \AA~(1.5--12.4 keV)~SXR flux and the $\textgreater$10 MeV proton flux, both observed simultaneously by the $Geostationary$ $Operational$ $Environmental$ $Satellite$ (\textit{GOES}) system. Although likely due to Big Flare Syndrome, correlations like this are still extremely useful, because they enable an empirical means of estimating particle fluxes from observations of photons -- essential for the study of distant stars.

Not all flares produce SEPs, and there may be a fundamental difference between SEP flares and non-SEP flares. \cite{Belov2005} note that an arbitrary SXR flare has a $\textless$0.4\% chance of being associated with a proton enhancement, although this estimate would likely increase if it were easier to confidently identify associated flares and SEPs, but the probability increases for larger flares. The \textit{GOES} flare classification scheme (A, B, C, M, X) is based solely on the peak 1--8 \AA~SXR flux as observed from Earth (1 AU), and each letter represents an increased order of magnitude from 10$^{-8}$ W m$^{-2}$ to 10$^{-3}$ W m$^{-2}$ (Table~\ref{table:GOES classification}). For example, a C3.5-class flare has a peak 3.5 $\times$~10$^{-6}$ W m$^{-2}$ SXR flux at 1 AU. Approximately 20\% of C-class and $\sim$100\% of X3-class flares occur with CMEs \citep{Yashiro2006}. Thus, estimating the \textit{GOES} flare classification of an observed stellar flare is important for estimating the probability of associated SEPs.

To estimate the SEP flux during the great AD Leo flare of 1985 \citep{Hawley1991}, \cite{Segura2010} used scaling relations for active M dwarfs between broadband near-UV and 1--8 \AA~flare flux \citep{Mitra-Kraev2005} and solar scaling relations between 1--8 \AA~flare flux and $\textgreater$10 MeV proton flux \citep{Belov2005,Belov2007}. Much of the recent UV flare data of M dwarfs has been confined to the far-UV \citep{Loyd2014,France2016}, where \textit{HST}'s COS and STIS spectrographs are most efficient, so we have developed a new method of particle flux estimation from far-UV emission line flares (Section~\ref{sec:uvproton}). 

Ideally, we would directly compare protons with a far-UV emission line directly accessible by \textit{HST}, but high-cadence, disk-integrated, spectrally-resolved far-UV observations of the Sun do not exist. To improve the method of SEP estimation from observed flares, we search for a potential correlation between energetic protons ($\textgreater$10 MeV) received at Earth and an extreme-UV emission line, \HeII~at 304 \AA, which has a similar formation temperature to two high-S/N lines observable in far-UV spectra (\SiIV~$\lambda \lambda$1393,1402~and \HeII~$\lambda$1640). 

This paper is organized as follows. In Section~\ref{sec:ObservationsReductions}, we describe our sample of M dwarfs and list the sources of the observations used in this work. We also describe the reductions performed. Section~\ref{sec:eqw} describes the method we used to measure the \cak~equivalent widths of our sample of M dwarfs, and in Section~\ref{sec:UVCaII_relation} we present the UV--\ca~scaling relations. In Section~\ref{sec:uvproton} we describe the UV--proton scaling relations and their application. We present a summary of the main findings of this work in Section~\ref{sec:Conclusions}.
\clearpage
\begin{deluxetable}{ccccccccccc}
\rotate
\tabletypesize{\scriptsize}
\tablecolumns{11}
\tablewidth{0pt}
\tablecaption{The M dwarf sample\label{table:targets}} 
\tablehead{\colhead{No.} &
                  \colhead{Star} & 
                  \colhead{$d$} & 
                  \colhead{$R$} & 
                  \colhead{Spectral} &
                  \colhead{$T_{\rm eff}$} &
                  \colhead{$P_{\rm rot}$} &
                  \colhead{W$_{\rm \lambda}$} (Ca II K) &
                  \colhead{W$_{\rm \lambda,corr}$} (Ca II K) &
                  \colhead{Data} &
                  \colhead{No. of} \\
                  \colhead{} & 
                  \colhead{} & 
                  \colhead{(pc)} & 
                  \colhead{($R_{\odot}$)} & 
                  \colhead{Type} &
                  \colhead{(K)} &
                  \colhead{(days)} &
                  \colhead{(\AA)} &
                  \colhead{(\AA)} &
                  \colhead{included$^{\dagger}$} &
                  \colhead{Ca II K~spectra}
                  }
\startdata
1 & GJ 832* & 5.0 $^a$ & 0.56 $^{f}$ & M2 $^{f}$ & 3590 $^{f}$ & 45.7 $^q$ & 0.88$\pm$0.09 & 0.73$\pm$0.07 & H, X, R & 8 \\
2 & GJ 876* & 4.7 $^a$ & 0.38 $^g$ & M4 $^g$ & 3130 $^g$ & 96.7 $^r$& 0.82$\pm$0.15 & 0.17$\pm$0.03 & H, X, R, D & 314\\
3 & GJ 581* & 6.2 $^a$ & 0.30 $^{z,h}$ & M2.5 $^{h}$ & 3500 $^{h}$ & 132.5 $^q$ & 0.36$\pm$0.08 & 0.25$\pm$0.05 & H & 351\\
4 & GJ 176* & 9.3 $^a$ & 0.45 $^g$ & M2.5 $^g$ & 3680 $^g$ & 39.5 $^s$ & 1.76$\pm$0.27 & 1.76$\pm$0.27 & H, D & 256 \\
5 & GJ 436* & 10.1 $^a$ & 0.45 $^{aa,i}$ & M3 $^{i}$ & 3420 $^{i}$ & 39.9 $^q$ & 0.58$\pm$0.07 & 0.33$\pm$0.04 & H, D & 257 \\
6 & GJ 667C* & 6.8 $^a$ & 0.46 $^j$ & M1.5 $^o$ & 3450 $^o$ & 103.9 $^q$ & 0.44$\pm$0.11 & 0.27$\pm$0.06 & H & 39 \\
7 & GJ 1214* & 14.6 $^b$ & 0.21 $^k$ & M4.5 $^o$ & 2820 $^o$ & $\textgreater$100 $^t$ & 1.16$\pm$0.2 & 0.040$\pm$0.007 & X, M & 10 \\
8 & AD Leo & 4.7 $^c$ & 0.44 $^{f}$  & M3 $^{f}$ & 3410 $^{f}$ & 2.2 $^u$ & 11.57$\pm$1.61 & 6.31$\pm$0.88 & H, R & 44 \\
9 & EV Lac & 5.1 $^a$ & 0.36 $^{f}$ & M4 $^{f}$ & 3330 $^{f}$ & 4.38 $^v$ & 14.86$\pm$2.52 & 6.47$\pm$1.10 & H & 24\\
10 & Proxima Cen* & 1.3 $^d$ & 0.14 $^l$ & M5.5 $^l$ & 3100 $^l$ & 83.5 $^w$ & 13.7$\pm$5.93 & 2.50$\pm$1.08 & X, M, R & 7\\
11 & AU Mic & 9.9 $^a$ & 0.83 $^m$ & PMS M1 $^{m,p}$ & 3650 $^{p}$ & 4.85 $^{x}$ & 12.13$\pm$2.17 & 11.44$\pm$2.05 & H, R & 52 \\
12 & YZ CMi & 6.0 $^a$ & 0.36 $^{f}$ & M4 $^{f}$ & 3200 $^{f}$ & 2.78 $^{y}$ & 21.23$\pm$4.67 & 5.92$\pm$1.30 & H, X & 20\\
13 & GJ 821 & 12.2 $^a$ & 0.37 $^{f}$ & M2 $^{f}$ & 3670 $^{f}$ & -- & 0.33$\pm$0.06 & 0.32$\pm$0.06 & H & 26 \\
14 & GJ 213 & 5.8 $^a$ & 0.25 $^{f}$ & M4 $^{f}$ & 3250 $^{f}$ & -- & 0.44$\pm$0.12 & 0.15$\pm$0.04 & H & 7\\
15 & GJ 1132* & 12.0 $^e$ & 0.21 $^n$ & M4 $^n$ & 3270 $^n$ & 125 $^n$ & 1.40$\pm$0.01 & 0.504$\pm$0.005 & M & 1\\
\enddata
\tablecomments{Rotation period ($P_{\rm rot}$) uncertainties are typically $\geq$10\% and effective temperature ($T_{\rm eff}$) uncertainties are typically $\pm$100--200 K unless determined from interferometry (then $\pm$20--100 K). We have rounded all $T_{\rm eff}$ values to the nearest 10 K.} 
\tablenotetext{*}{Stars with known exoplanets.}
\tablenotetext{\dagger}{(H) Keck/HIRES, (X) VLT/XSHOOTER, (R) CASLEO/REOSC, (D) APO/DIS, (M) Magellan/MIKE.}
\tablerefs{(a) \citealt{VanLeeuwen2007}, (b) \citealt{Anglada-Escude2013}, (c) \citealt{Jenkins1952}, (d) \citealt{Lurie2014}, (e) \citealt{Jao2005}, (f) \citealt{Houdebine2016}, (g) **\citealt{VonBraun2014}, (h) **\citealt{VonBraun2011}, (i) **\citealt{VonBraun2012},  (j) \citealt{Kraus2011},  (k) \citealt{Berta2011}, (l) **\citealt{Demory2009}, (m) **\citealt{White2015}, (n) \citealt{Berta-Thompson2015a}, (o) \citealt{Neves2014}, (p) \citealt{McCarthy2012}, (q) \citealt{SuarezMascareno2015}, (r) \citealt{Rivera2005}, (s) \citealt{Robertson2015}, (t) \citealt{Newton2016a}, (u) \citealt{Hunt-Walker2012}, (v) \citealt{Roizman1984}, (w) \citealt{Benedict1998}, (x) \citealt{Vogt1983}, (y) \citealt{Pettersen1980}, (z) \citealt{Henry1994}, (aa) \citealt{Hawley1996}. **Interferometry measurements.} 
\end{deluxetable}
\clearpage

\section{Observations \& Reductions} \label{sec:ObservationsReductions}

\subsection{The M dwarf sample}

We selected stars with \textit{HST} UV spectra and ground-based optical spectra either obtained directly by us or available in the VLT/XSHOOTER or Keck/HIRES public archives. The 15 stars that meet this criteria (listed in Table~\ref{table:targets}) are all early-to-mid M dwarfs, nearby ($d$ $\textless$~15 pc), and exhibit a broad range of rotation periods (2 to $\textgreater$100 days) and a broad range of ages ($\sim$10 Myr to $\sim$10 Gyr). Nine of the 15 M dwarfs are known to host exoplanets. 

Seven of the 15 stars are exoplanet host stars from the MUSCLES Treasury Survey (stars 1--7 in Table~\ref{table:targets}), and are weakly-active with H$\alpha$~absorption spectra and rotation periods greater than 39 days. These stars are all likely a few billion years old, and they range from M1.5--M4.5 spectral type. We also included other weakly-active M dwarfs, including MEarth planet host GJ 1132 \citep{Berta-Thompson2015} and two stars from the ``Living with a Red Dwarf" program (stars 13--15; \citealt{Guinan2016}). To increase the diversity in our sample, we included the well-known ``flare" stars with \textit{HST} observations (stars 8--12 in Table~\ref{table:targets}). These stars are highly active (H$\alpha$~emission spectra), have short rotation periods ($\textless$~7 days), and are likely young ($\textless$~1 Gyr), with the exception of Proxima Centauri ($P_{\rm rot}$ = 83.5 days, $\sim$5 Gyr old; see \citealt{Reiners2008,Davenport2016}).

\subsection{M Dwarf UV and optical data} \label{sec:UVopticaldata}

A goal of the MUSCLES Treasury Survey was to obtain ground-based optical spectra contemporaneous with the \textit{HST} UV observations. Scheduling changes and weather did not allow for any truly simultaneous UV--optical observations, but several targets have spectroscopic data obtained with the Dual Imaging Spectrograph (DIS) on the ARC 3.5m telescope at Apache Point Observatory (APO) or the REOSC echelle spectrograph on the 2.15m telescope at Complejo Astron\'omico El Leoncito (CASLEO) within a day or two of the \textit{HST} observations. We also gathered spectra of GJ1132, GJ1214, and Proxima Cen on the nights of 2016 March 7--9 using the MIKE echelle spectrograph on the Magellan Clay telescope. Because M dwarfs are prime targets of radial velocity exoplanet searches, there is a wealth of high-resolution \ca~spectra in the public archives of major observatories, including VLT and Keck. The archival spectra comprise the bulk of our \ca~measurements. 

APO/DIS ($R$~$\sim$~2500) and CASLEO/REOSC ($R$~$\sim$~12,000) spectra were reduced using standard IRAF\footnote{IRAF is distributed by the National Optical Astronomy Observatory, which is operated by the Association of Universities for Research in Astronomy, Inc. under cooperative agreement with the National Science Foundation.} routines. See details in \citealt{Cincunegui2004} for CASLEO reductions and \cite{Cincunegui2007} and \cite{Buccino2014} for presentation of some of the Proxima Cen and AD Leo observations. Magellan/MIKE spectra ($R$~$\sim$~25,000) were reduced using the standard MIKE pipeline included in the Carnegie Python Distribution (CarPy). Science-level VLT/XSHOOTER ($R$~$\sim$~6000) data products were obtained from the ESO Science Archive Facility, and Keck/HIRES ($R$~$\sim$~60,000) pipeline-reduced spectra were obtained from the KOA archive. As we are interested in looking at as many spectra as possible and measuring only equivalent widths of the \cak~line, the pipeline-extracted spectra using \texttt{MAKEE}\footnote{http://www.astro.caltech.edu/$\sim$tb/makee/} suit our purposes well. We did not use spectra where the automatic extraction failed or the wavelength calibration was incorrect. We also did not coadd the adjacent orders on which \cak~appears, but instead averaged the equivalent width measurements (Section~\ref{sec:eqw}) from each order to create one equivalent width measurement per echellogram.

The \textit{HST} UV spectra were obtained either from the MAST online archive, including the MUSCLES High-Level Science Products\footnote{https://archive.stsci.edu/prepds/muscles/} (HLSPs; \citealt{Loyd2016}), or the StarCAT portal\footnote{http://casa.colorado.edu/$\sim$ayres/StarCAT/}. GJ 1132's (star 15) STIS G140M spectrum was obtained on 2016 February 13. For the seven MUSCLES M dwarfs (stars 1--7), the UV line fluxes come from \cite{Youngblood2016} and \cite{France2016}. The UV emission lines of stars 8--14 were directly measured from archival \textit{HST} spectra or obtained from \cite{Wood2005}. 

For AD Leo (star 8) and Proxima Centauri (star 10), we reconstructed the \Lya~profiles using the methods described in \cite{Youngblood2016}. Proxima Cen's reconstruction is included as part of a 5 \AA--5.5 $\mu$m SED on the MUSCLES HLSP website\footnote{https://archive.stsci.edu/prepds/muscles/}. The differences between our \Lya~reconstructions and those presented in \cite{Wood2005} are small, $\sim$20\% in integrated \Lya~flux for AD Leo and $\sim$4\% in integrated \Lya~flux for Proxima Cen. We also reconstructed the \Lya~profile of GJ 1132 (star 15; $F$(\Lya) = (2.64$^{+2.58}_{-0.65}$)~$\times$~10$^{-14}$ erg cm$^{-2}$ s$^{-1}$) for the first time using the methods from \cite{Youngblood2016}. Note that the \Lya~error bars have been averaged for symmetry in Table~\ref{table:UVfluxes1}. All UV fluxes used in this work are printed in Tables~\ref{table:UVfluxes1} and \ref{table:UVfluxes2}.

\subsubsection{Interstellar medium corrections} \label{sec:ISM}
5/9 of our UV emission lines (\Lya, \MgII, \CII, \SiII, and \SiIII) are affected by absorption from the interstellar medium (ISM), and here we describe our attempts to mitigate the effect of ISM absorption on our measurements. All the reported \Lya~fluxes have been reconstructed from the wings of the observed line profile using the technique described in \cite{Youngblood2016} or \cite{Wood2005} for EV Lac. The \MgII~fluxes were corrected uniformly for a 30\% ISM absorption, assuming a typical log$_{10}$ $N$(\MgII) $\sim$~13 for stars within 20 pc \citep{Redfield2002}. Many of the \MgII~observations are not sufficiently resolved to allow for a profile reconstruction, and we do not apply a correction factor scaling either with distance or \HI~column densities measured along the line-of-sight, because the Mg$^+$ abundance varies in the local ISM. \CII~1334 \AA~is the most significantly-impacted of the \CII~$\lambda \lambda$1334, 1335 doublet, and so we excluded its contribution from the reported \CII~fluxes. 

We do not apply ISM corrections to the \SiII~(see \citealt{Redfield2004} for a discussion of \SiII~absorption in the local ISM)~or \SiIII~fluxes, noting that the ISM's effect on \SiIII~is likely small. The intrinsic narrowness of the M dwarf emission lines may mean that for some sightlines, the ISM absorption coincides with the stellar emission line, but for others the ISM absorption is shifted away from the emission, and the line's flux is not attenuated. Ca$^+$~from the ISM can significantly attenuate \ca~H \& K, but only for distant stars ($d$ $\textgreater$~100 pc; \citealt{Fossati2017}). Because all of our targets are within 15 pc, \ca~ISM absorption is likely negligible, and we do not apply a correction.

\clearpage
\begin{deluxetable}{cccccccccccc}
\rotate
\tabletypesize{\scriptsize}
\tablecolumns{12}
\tablewidth{0pt}
\tablecaption{UV emission line fluxes of the M dwarf sample (continued to Table~\ref{table:UVfluxes2}) \label{table:UVfluxes1}} 
\tablehead{\colhead{No.} &
                  \colhead{Star} & 
                  \colhead{Si III} & 
                  \colhead{$\sigma_{\rm Si III}$} &
                  \colhead{Ly$\alpha$ $^a$} & 
                  \colhead{$\sigma_{\rm Ly\alpha}$ $^a$} &
                  \colhead{Si II} &
                  \colhead{$\sigma_{\rm Si II}$} &
                  \colhead{C II} &
                  \colhead{$\sigma_{\rm C II}$} &
                  \colhead{Mg II} &
                  \colhead{$\sigma_{\rm Mg II}$}
                  }
\startdata
1 & GJ 832* & 2.55E-15  & 6.33E-17 & 9.50E-13 & 6.00E-14 & 7.65E-16 & 6.33E-17 & 3.78E-15 & 8.18E-17 & 1.84E-13 & 2.02E-15\\
2 & GJ 876* & 8.11E-15 & 1.01E-16 & 3.90E-13 & 4.00E-14 & 1.19E-15 & 7.06E-17 & 1.06E-14 & 1.26E-16 & 3.11E-14 & 9.31E-16 \\
3 & GJ 581* & 2.94E-16 & 3.54E-17 & 1.10E-13 & 3.00E-14 & 7.93E-17 & 4.41E-17 & 4.81E-16 & 4.26E-17 & 1.88E-14 & 7.54E-16  \\
4 & GJ 176* &  2.15E-15 & 5.84E-17 & 3.90E-13 & 2.00E-14 & 7.34E-16 & 6.49E-17	 & 5.43E-15 & 9.66E-17	 & 1.56E-13 & 1.53E-15 \\
5 & GJ 436* &  5.15E-16 & 4.00E-17 & 2.10E-13 & 3.00E-14 & 1.87E-16 & 4.47E-17	 & 1.09E-15 & 5.85E-17	 & 3.84E-14 & 1.01E-15 \\
6 & GJ 667C* & 5.07E-16 & 3.91E-17 & 5.20E-13 & 9.00E-14	 & 1.29E-16 & 6.13E-17 & 6.50E-16 & 4.63E-17 & 4.12E-14 & 1.10E-15\\
7 & GJ 1214* &  8.05E-17 & 2.77E-17 & 1.30E-14 & 1.00E-14 & 1.36E-17 & 2.09E-17 & 9.83E-17 & 3.00E-17	 & 1.66E-15 & 2.70E-16 \\
8 & AD Leo &  1.41E-13 & 8.88E-16 & 8.07E-12 & 2.00E-13 & 1.64E-14 & 2.58E-16 & 1.66E-13 & 4.83E-16 & 2.79E-12 $^b$ & 2.79E-13 $^b$ \\
9 & EV Lac & 3.86E-14 & 1.46E-15 & 2.75E-12 $^b$ & 5.50E-13 $^b$  & 2.49E-15 & 4.95E-16 & 3.76E-14 & 6.42E-16 & -- & -- \\
10 & Proxima Cen* & 2.03E-14 & 7.57E-16 & 4.37E-12 & 7.00E-14 & 2.14E-15 & 2.08E-16 & 2.61E-14 & 2.79E-16 & -- & -- \\
11 & AU Mic & 8.78E-14 & 3.45E-16 & 1.07E-11 & 4.00E-13 & 1.09E-14 & 8.97E-17 & 1.44E-13 & 1.81E-16 & 4.20E-12 $^b$ & 4.20E-13 $^b$ \\
12 & YZ CMi & -- & -- & -- & -- & -- & -- & -- & -- & 1.10E-12 & 1.42E-15 \\
13 & GJ 821 & 5.58E-16 & 6.56E-16 & -- & -- & 9.52E-17 & 9.52E-18 & 1.05E-16 & 2.89E-17 & 2.62E-14 & 2.69E-15 \\
14 & GJ 213 & -- & -- & -- & -- & 1.62E-16 & 1.62E-17 & 2.32E-16 & 2.32E-17 & 7.13E-15 & 2.30E-15 \\
15 & GJ 1132* & -- & -- & 2.64E-14 & 1.62E-14 & -- & -- & -- & -- & -- & -- \\
\enddata
\tablecomments{All emission line fluxes are as observed from Earth (erg cm$^{-2}$ s$^{-1}$). All flux measurements for stars 1--7 come from \cite{Youngblood2016} and \cite{France2016}. The \Lya~reconstructed flux for AU Mic also comes from \cite{Youngblood2016}. All other measurements are from this work, except where noted.}
\tablenotetext{*}{Stars with known exoplanets.}
\tablenotetext{a}{Reconstructed \Lya~fluxes. Asymmetric error bars from \cite{Youngblood2016} have been averaged.}
\tablenotetext{b}{\cite{Wood2005}.}
\end{deluxetable}
\clearpage

\clearpage
\begin{deluxetable}{ccccccccccc}
\rotate
\tabletypesize{\scriptsize}
\tablecolumns{11}
\tablewidth{0pt}
\tablecaption{UV emission line fluxes of the M dwarf sample -- continuation of Table~\ref{table:UVfluxes1} \label{table:UVfluxes2}} 
\tablehead{\colhead{No.} &
                  \colhead{Star} & 
                  \colhead{Si IV} &
                  \colhead{$\sigma_{\rm Si IV}$} &
                  \colhead{He II} &
                  \colhead{$\sigma_{\rm He II}$} &
                  \colhead{C IV} &
                  \colhead{$\sigma_{\rm C IV}$} &
                  \colhead{N V} & 
                  \colhead{$\sigma_{\rm N V}$} &
                  \colhead{EUV $^c$}
                  }
\startdata
1 & GJ 832* & 3.33E-15 & 9.19E-17 & 6.81E-15 & 2.93E-16 & 7.64E-15 & 2.88E-16 & 3.51E-15 & 7.54E-17 & 8.65E-13\\
2 & GJ 876* & 8.44E-15 & 1.36E-16 & 5.51E-15 & 2.92E-16 & 2.29E-14 & 4.34E-16 & 1.07E-14 & 1.26E-16 & 3.43E-13\\
3 & GJ 581* & 4.44E-16 & 6.71E-17 & 7.22E-16 & 1.90E-16 & 1.84E-15 & 1.84E-16 & 5.34E-16 & 4.48E-17 & 1.01E-13 \\
4 & GJ 176* & 2.30E-15 & 8.64E-17 & 6.73E-15 & 3.03E-16 & 1.23E-14 & 2.83E-16 & 3.10E-15 & 7.56E-17 & 3.57E-13\\
5 & GJ 436* & 6.81E-16 & 6.92E-17 & 1.50E-15 & 2.13E-16 & 2.39E-15 & 1.76E-16 & 9.56E-16 & 5.04E-17 & 1.90E-13\\
6 & GJ 667C* & 8.25E-16 & 7.17E-17 & 1.94E-15 & 2.25E-16 & 2.97E-15 & 2.05E-16 & 7.16E-16 & 5.31E-17 & 4.73E-13\\
7 & GJ 1214* & 4.79E-17 & 3.74E-17 & 7.43E-17 & 1.44E-16 & 5.23E-16 & 1.21E-16 & 1.84E-16 & 3.58E-17 & 1.29E-14\\
8 & AD Leo & 1.74E-13 & 5.39E-16 & 3.02E-13 & 9.45E-16 & 5.66E-13 & 1.21E-15 & 1.41E-13 & 5.89E-16 & 8.16E-12\\
9 & EV Lac & 1.82E-14 & 9.76E-16 & 7.41E-14 & 1.58E-15 & 1.41E-13 & 1.96E-15 & 3.91E-14 & 9.03E-16 & 2.60E-12 \\
10 & Proxima Cen* & 2.72E-14 & 3.83E-16 & 3.08E-14 & 5.87E-16 & 1.38E-13 & 9.13E-16 & 4.00E-14 & 4.56E-16 & 3.86E-12\\
11 & AU Mic & 9.53E-14 & 1.63E-16 & 2.75E-13 & 3.50E-16 & 3.55E-13 & 3.85E-16 & 7.42E-14 & 1.82E-16 & 1.42E-11 \\
12 & YZ CMi & -- & -- & -- & -- & -- & -- & -- & -- & -- \\
13 & GJ 821 & 8.16E-17 & 4.29E-17 & 6.84E-16 & 2.53E-16 & 6.69E-16 & 1.90E-16 & 1.95E-16 & 1.95E-17 & -- \\
14 & GJ 213 & 2.13E-16 & 2.13E-17 & 2.57E-16 & 2.57E-17 & 5.22E-16 & 1.55E-16 & 3.58E-16 & 3.58E-17 & -- \\
15 & GJ 1132* & -- & -- & -- & -- & -- & -- & -- & -- & 2.31E-14 \\
\enddata
\tablecomments{All emission line fluxes are as observed from Earth (erg cm$^{-2}$ s$^{-1}$). All flux measurements for stars 1--7 come from \cite{Youngblood2016} and \cite{France2016}. All other measurements are from this work.}
\tablenotetext{*}{Stars with known exoplanets.}
\tablenotetext{c}{Calculated using the reconstructed \Lya~fluxes (Table~\ref{table:UVfluxes1}) and scaling relations from \cite{Linsky2014}, which have been reprinted in Table~\ref{table:CaK_EUV}.}
\end{deluxetable}
\clearpage

\subsection{Solar X-ray, UV, and proton data}

We utilize time-series solar irradiance (disk-integrated) measurements from the Extreme ultraviolet Variability Experiment (EVE) suite of instruments onboard the \textit{Solar Dynamics Observatory} (\textit{SDO}; \citealt{Woods2012}). We use the \HeII~(304 \AA) irradiance from the 2010--2014 era from the MEGS-A channel (50--370 \AA~with 1 \AA~spectral resolution). MEGS-A operates at a 10-second cadence, but we use one-minute averages in this work to reduce noise.

Time-series solar SXR irradiance measurements and \textit{in situ} proton measurements come from the \textit{Geostationary Operational Environmental Satellite} (\textit{GOES}) system\footnote{https://www.ngdc.noaa.gov/stp/satellite/goes/index.html}. SXRs are measured in a 1--8 \AA~(1.5--12.4 keV) band. Note that we do not divide the \textit{GOES} 1--8 \AA~flux by 0.7, the recommended correction factor, to obtain absolute flux units. This is because we utilize the \textit{GOES} SXR flare classification scheme below, which operates on data that have not been corrected, and recent SXR observations with the MinXSS cubesat suggest this correction factor is incorrect \citep{Woods2017}. We utilize the proton measurements from the $\textgreater$10 MeV and $\textgreater$30 MeV channels.

\section{\cak~Equivalent Widths} \label{sec:eqw}

To isolate the chromospheric \cak~emission line so we can compare it to the UV emission line fluxes, we must correct for the absorption by subtracting a radiative equilibrium model. This is particularly important for the low resolution spectra, which typically cannot resolve the emission cores. Only the CASLEO/REOSC spectra are flux calibrated, so to be consistent with the rest of the spectra, we normalized the flux in the isolated \cak~emission core to nearby continuum, making this measurement akin to an equivalent width. \cite{Busa2007}, \cite{Marsden2009}, and \cite{Walkowicz2009} used this ``residual" equivalent width technique in their analyses of \ca~lines. Note that we do not use the widely-used S-index \citep{Wilson1978,Vaughan1978} or $R'_{\rm HK}$ index (e.g., \citealt{Noyes1984}), because for our medium-resolution spectra the H$\epsilon$ emission line is blended with the \cah~emission core line.

\begin{deluxetable}{cc|cc}
\tablecolumns{4}
\tablewidth{0pt}
\tablecaption{Correction factors for calculating $W_{\rm \lambda, corr}$ (Equation~\ref{eq:corrfactor})  \label{table:spline}} 
\tablehead{\colhead{$T$} & 
                  \colhead{$x$($T$)/$x$(3680 K)} & 
                  \colhead{$T$} & 
                  \colhead{$x$($T$)/$x$(3680 K)}
                  }
\startdata
2400 & 1.1613$\times$10$^{-3}$ & 3300 & 0.39680\\
2500 & 2.5175$\times$10$^{-3}$ & 3400 & 0.53075 \\
2600 & 5.6663$\times$10$^{-3}$ & 3500 & 0.68191 \\
2700 & 1.2907$\times$10$^{-2}$ & 3600 & 0.85061 \\
2800 & 2.9381$\times$10$^{-2}$ & 3700 & 1.0425 \\
2900 & 5.9956$\times$10$^{-2}$ & 3800 & 1.2642\\
3000 & 0.11061 & 3900 & 1.5233\\
3100 & 0.18269 & 4000 & 1.8277\\
3200 & 0.27899 \\
\enddata
\tablecomments{To calculate the correction factors $x$($T$)/$x$(3680 K), we computed $x$($T$), the average stellar surface flux values from 3936.9--3939.9 \AA~for PHOENIX models from the \cite{Husser2013} grid for a range of temperatures $T$ (K), all with log$_{10}$ $g$ = 5 and [Fe/H] = 0 (solar). From a cubic spline fit to the $x$($T$) values, we find $x$(3680 K) = 7.7453$\times$10$^{12}$ erg cm$^{-2}$ s$^{-1}$ cm$^{-1}$.} 
\end{deluxetable}

\begin{figure}[t]
   \begin{center}
   
     \subfigure{
          \includegraphics[width=0.6\textwidth]{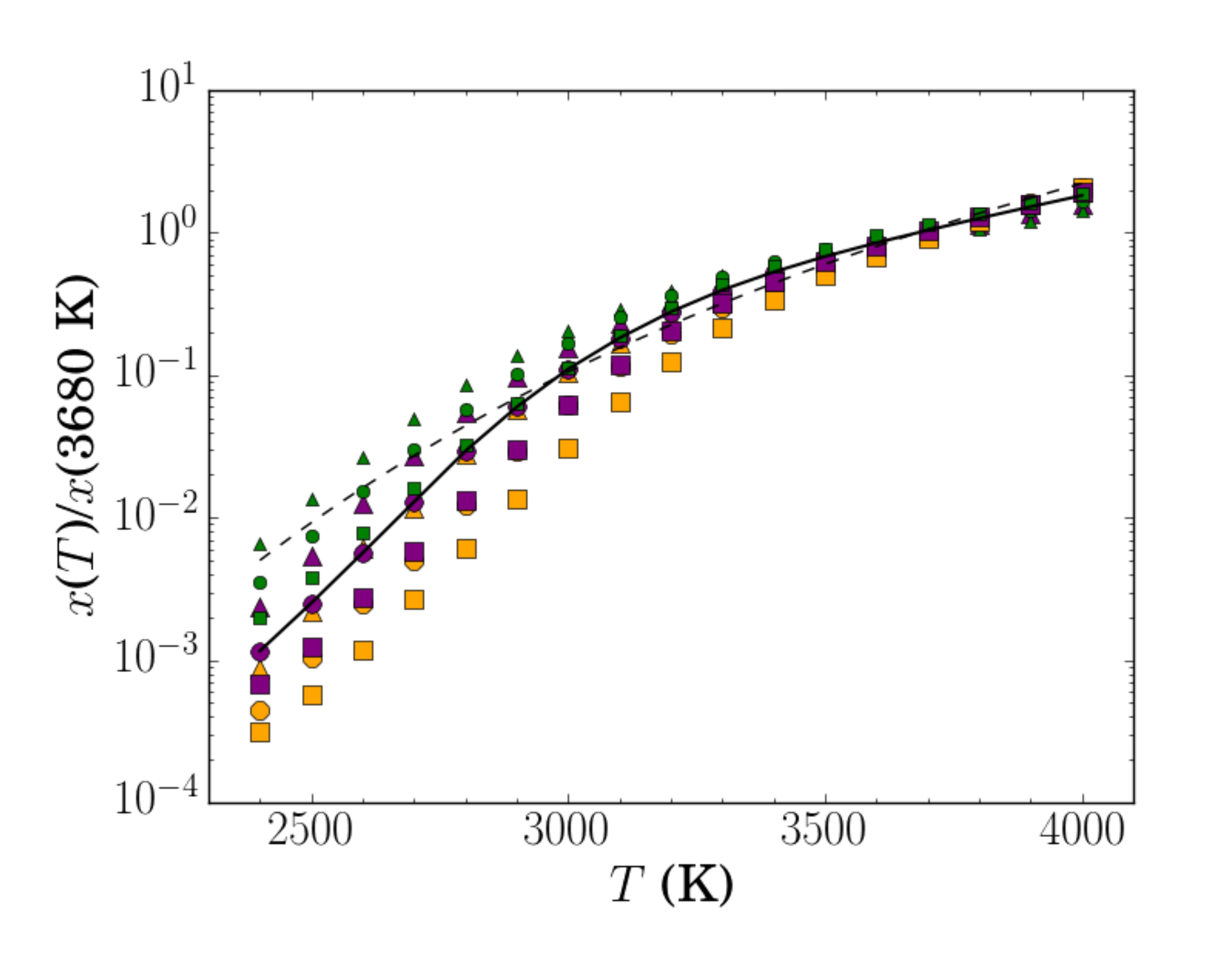}
          }

   \end{center}
    \caption{The correction factor $x$($T$)/$x$(3680 K) used for calculating $W_{\rm \lambda, corr}$ from $W_{\lambda}$ as a function of stellar effective temperature $T$ (Equation~\ref{eq:corrfactor}). The solid black curve shows the cubic spline fit to the purple circles, which correspond to the data in Table~\ref{table:spline}. This figure shows the spread in correction factor values when varying metallicity and surface gravity. Purple represents [Fe/H] = 0, orange [Fe/H] = -0.5, and green [Fe/H] = +0.5. The circles represent log$_{10}$ $g$ = 5.0, triangles log$_{10}$ $g$ = 4.5, and squares log$_{10}$ $g$ = 5.5. The dashed black curve shows a correction factor based on a blackbody curve.
        }
    \label{fig:corr_factor}

\end{figure}  

The residual equivalent width $W_{\lambda}$ is given by:

\begin{equation}
W_{\rm \lambda}  = \frac{\int_{\lambda_1}^{\lambda_2} F_{\rm observed}\,d\lambda -  \int_{\lambda_1}^{\lambda_2} F_{\rm PHOENIX}\,d\lambda}{\overline{F}_{\rm continuum}}.
\label{eq:eqw1}
\end{equation}

\noindent $F_{\rm observed}$ is the observed \cak~profile, $F_{\rm PHOENIX}$ is the radiative equilibrium PHOENIX model \citep{Husser2013} scaled to $F_{\rm observed}$ in the \cak~absorption wings, $\lambda_1$ and $\lambda_2$ are the interactively-chosen bounds of integration around the narrow chromospheric \cak~emission, and $\overline{F}_{\rm continuum}$ is the average observed flux in the continuum region (before subtraction of the PHOENIX model) from 3937--3940 \AA. We selected the 3937--3940 \AA~region for continuum normalization because it is close to the \cak~line (3933 \AA), and the spectral response curve of the various CCDs we are using are unlikely to vary significantly over $\sim$7 \AA. \cite{Walkowicz2009} and \cite{Rauscher2006} normalize their effective equivalent widths using continuum regions 3952.8--3956.0 \AA~and 3974.8--3976.0 \AA. We do not believe there is a significant difference between these continuum choices, because neither are truly continuum due to the high density of absorption lines in this spectral region.

The PHOENIX models were retrieved from the \cite{Husser2013} grid using literature values for $T_{\rm eff}$ (Table~\ref{table:targets}) and assuming log$_{10}~g$ = 5 and [Fe/H] = 0 (solar). However, for the 7 MUSCLES targets, we use the PHOENIX models incorporated into the high-level science products available on MAST. See \cite{Loyd2016} for details about the parameters used to retrieve these models. The PHOENIX models have resolution $R$ = 500,000 around \cak~\citep{Husser2013} and were convolved with a Gaussian kernel to match the resolution of the spectra from the various instruments used (Section~\ref{sec:ObservationsReductions}). The PHOENIX models were also shifted in velocity space to the rest frame of each target before fitting. We scaled the PHOENIX spectra to match 3 points in the broad \cak~absorption wings.

The M dwarfs in our sample have a broad range of effective temperatures, so we must account for a dependence on spectral type in this residual equivalent width measurement (see Appendix~\ref{app: Though Experiment} for more details on the following description). Stars of earlier spectral type have brighter continuum fluxes than later type stars, affecting the normalization of the emission core's flux. \cite{Walkowicz2009} restricted their \ca~analysis to a single spectral subtype (M3 V) to avoid the issue of a spectral type dependence. To correct the residual equivalent widths ($W_{\rm \lambda}$ from Equation~\ref{eq:eqw1}) for spectral type dependence, we normalized to the $W_{\rm \lambda}$ value for a reference star parameterized by $T_{\rm eff}$~= 3680 K, log $g$ = 5, and [Fe/H] = 0. The normalization or ``correction" factor is the ratio of the star's average continuum flux and the reference star's average continuum flux. Both continuum averages come from the PHOENIX models. The corrected residual equivalent width is given by

\begin{equation}
W_{\rm \lambda, corr} = W_{\lambda}\,\times \,\frac{x(T)}{x({\rm 3680 K})}.
\label{eq:corrfactor}
\end{equation}

\noindent $x$($T$) is the average continuum flux value from 3936.9--3939.9 \AA~from a PHOENIX model for a star with temperature \textit{T}, and $x$(\rm 3680 K) is the PHOENIX model's average continuum flux value for the reference star. The reference star's effective temperature corresponds to GJ 176 and was chosen arbitrarily. To find the continuum values, $x(T)$, we used a grid of PHOENIX models from $T_{\rm eff}$ = 2400 K to $T_{\rm eff}$ = 4000 K, all with log$_{10}$~$g$ = 5 and [Fe/H] = 0 (solar). We fit a cubic spline function to the average flux values in the 3936.9--3939.9 \AA~region to determine the correction factor $x(T)$/$x({\rm 3680 K})$ (Table~\ref{table:spline}, Figure~\ref{fig:corr_factor}). The uncertainty in $W_{\rm \lambda, corr}$ introduced by the correction factor has two roughly equivalent sources: the uncertainty in $T_{\rm eff}$ and the assumption of log$_{10}$~$g$ = 5 and [Fe/H] = 0 for all our stars. Uncertainties in $T_{\rm eff}$ are likely 100--200 K, and this translates to a $\sim$20\% uncertainty in the correction factor at the high temperature end ($T_{\rm eff}$ = 3700 K), to $\sim$70\% uncertainty around $T_{\rm eff}$ = 3000 K, and $\sim$100\% uncertainty at the low temperature end ($T_{\rm eff}$ = 2400 K). The surface gravity of our M dwarf sample ranges from $\sim$4.75--5.0, and given the coarseness of the PHOENIX grids, log$_{10}$~$g$ = 5 is a good assumption. The metallicity ranges from -0.5 to 0.5, but [Fe/H] = 0 is valid for most of the M dwarfs in our sample. Examining the average flux values in the 3936.9--3939.9 \AA~region for PHOENIX models sampling a range of surface gravity and metallicity values, we find that the dispersion in $x$($T$)/$x$(3680 K) values for a given temperature is of similar magnitude to the dispersion in $x$($T$)/$x$(3680 K) values (assuming log$_{10}$~g = 5 and [Fe/H] = 0) between $\Delta T_{\rm eff}$ = 200 K bins (Figure~\ref{fig:corr_factor}).

\begin{figure*}
   \begin{center}
   
     \subfigure{
          \includegraphics[width=\textwidth]{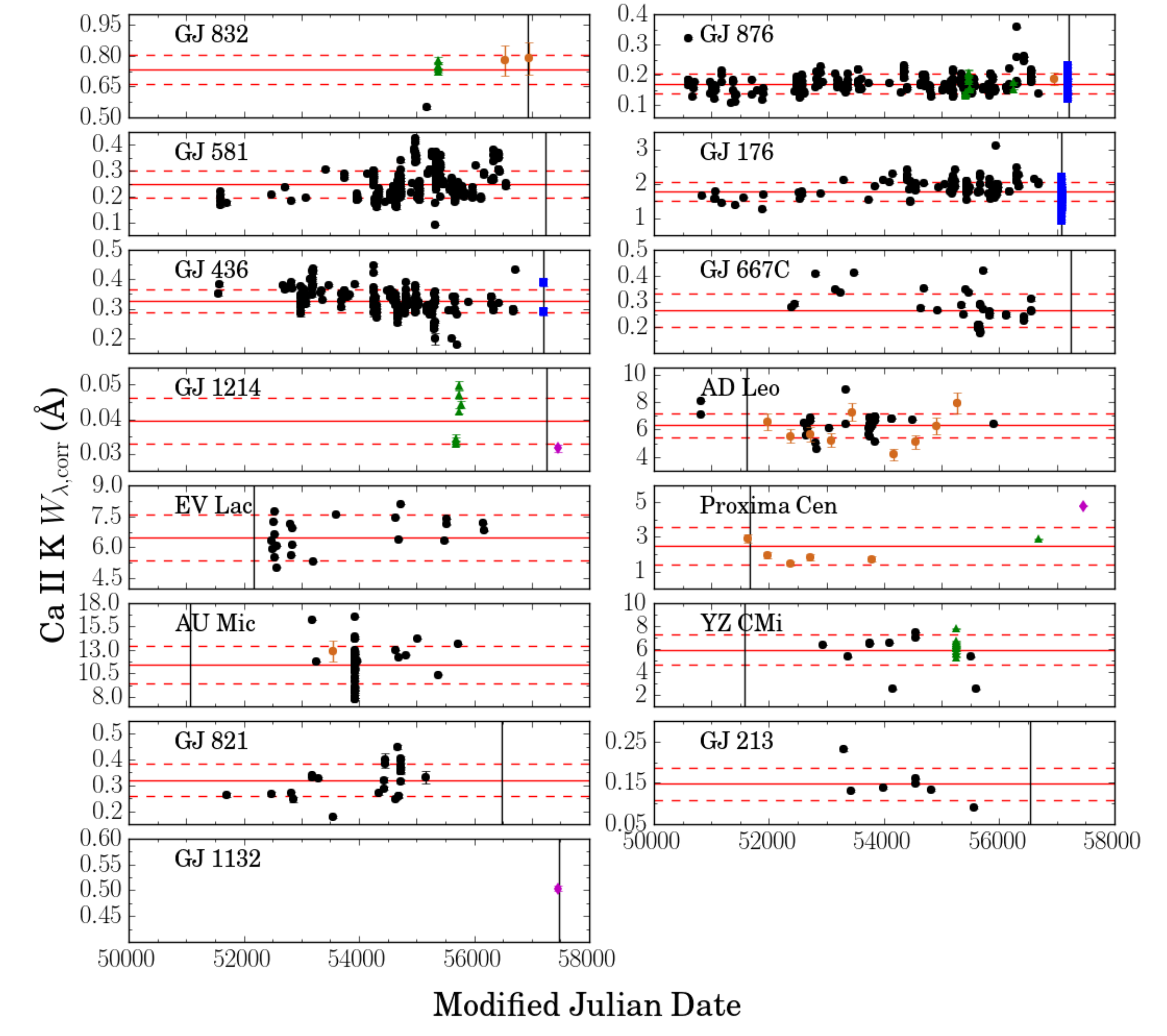}
          }

   \end{center}
    \caption{\cak~equivalent width ($W_{\rm \lambda, corr}$) lightcurves for our M dwarf sample displayed over the time period 1995 October 10 to 2017 September 4 (Modified Julian Date = 50,000 to 58,000). The solid horizontal red line shows the mean equivalent width, and the two dashed red lines show the standard deviation of the data. The vertical solid black lines show the dates of the UV observations used in this work (see Section~\ref{sec:correlations} and Tables~\ref{table:UVfluxes1} and \ref{table:UVfluxes2}). The black circles represent data from Keck/HIRES, green triangles are VLT/XSHOOTER, orange circles are CASLEO/REOSC, magenta diamonds are Magellan/MIKE, and blue squares are APO/DIS. 1--$\sigma$~photometric error bars are shown, but they are typically smaller than the data points.
        }
    \label{fig:time_eqw}

\end{figure*}  

\begin{figure}[t]
   \begin{center}
   
     \subfigure{
          \includegraphics[width=\textwidth]{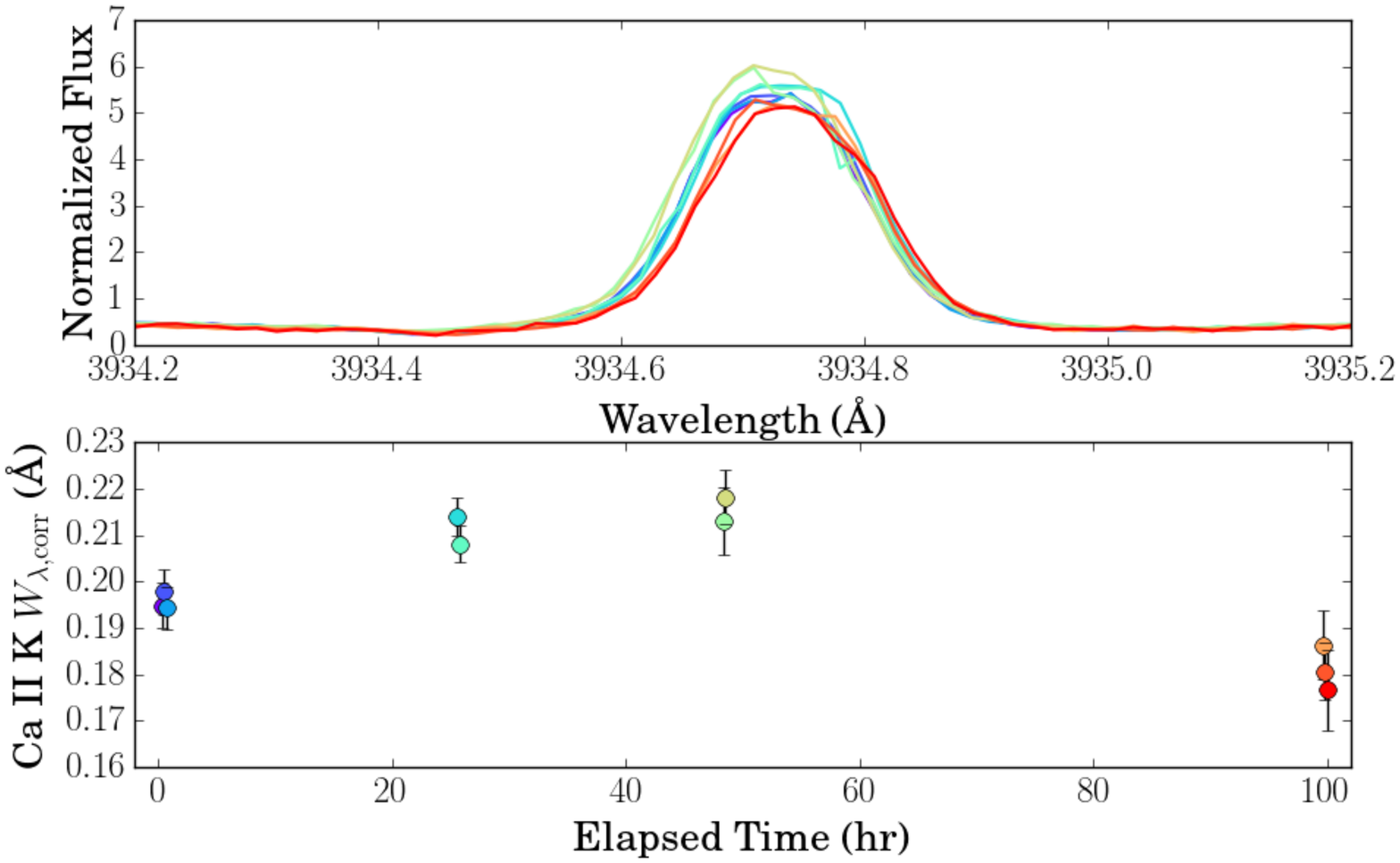}
          }

   \end{center}
    \caption{\textit{Top panel:} Normalized line profiles of the \cak~line for GJ 876 over the course of 3 days as observed by Keck/HIRES 2013 September 13--17. The apparent redshift in the line centroid is likely an artifact from the \texttt{MAKEE} automated wavelength calibration. \textit{Bottom panel:} \cak~corrected equivalent widths over the course of the 4 days. The colors correspond to the line colors in the top panel.
        }
    \label{fig:ca_profile_overplot}

\end{figure}  

We also examined the effect of PHOENIX model parameters on $W_{\rm \lambda, corr}$ during the model subtraction and the effect of interactively choosing the integration limits. Using a single Keck/HIRES spectrum and three PHOENIX model spectra, we measured $W_{\rm \lambda, corr}$ 30 times and compared to the ``true" measurement for that spectrum and its photometric error bar (the true measurements are shown in Figure~\ref{fig:time_eqw} and used to calculate the equivalent width values reported in Table~\ref{table:targets}). We find that the PHOENIX model chosen for subtraction has a negligible effect on the equivalent width measurements, but that the uncertainty introduced from differences in the interactively-chosen integration bounds is slightly larger than the photometric uncertainty. We do not correct the photometric error bars on the individual equivalent width measurements, and the error bars reported in Table~\ref{table:targets} come from the dispersion in the many equivalent width measurements made for each target (described fully in Section~\ref{sec:variability}).

\section{UV--\ca~Relation} \label{sec:UVCaII_relation}

In this section we present scaling relations between \cak~and far- and near-UV emission lines, as well as the total extreme-UV flux (100--912 \AA). See Appendix~\ref{app:uvuv} for a presentation of scaling relations between the far- and near-UV emission lines themselves.

\subsection{\cak~variability} \label{sec:variability}
\begin{figure}[t]
   \begin{center}
   
     \subfigure{
          \includegraphics[width=\textwidth]{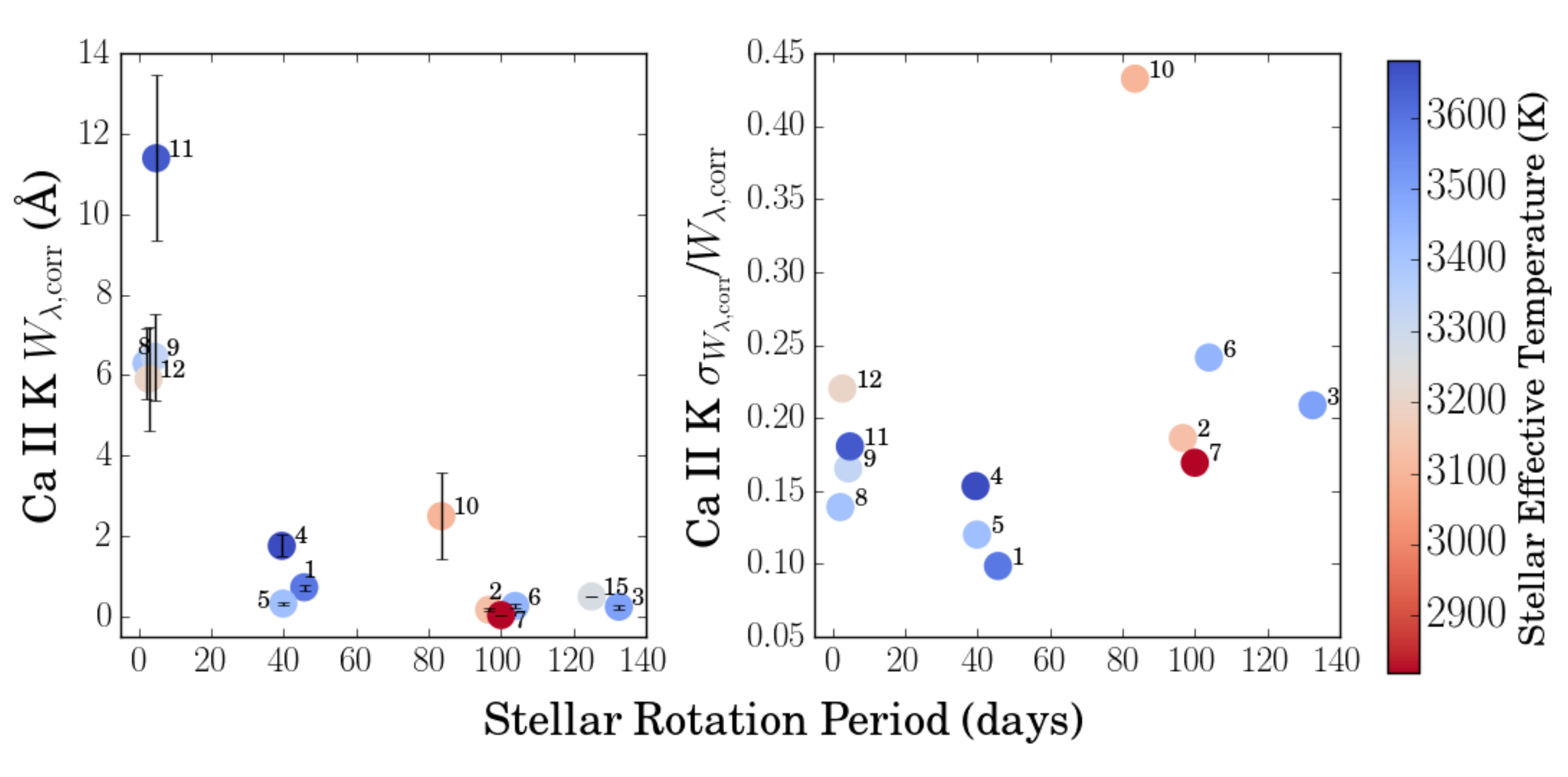}
          }

   \end{center}
    \caption{The stellar rotation period in days compared to the corrected \cak~equivalent widths (left panel) and the normalized scatter in the equivalent widths (right panel). Note that GJ 1132 (star 15) is not included in the right panel, because it only has one \cak~measurements, and GJ 821 (star 13) and GJ 213 (star 14) are not included in either panel, because they do not have $P_{\rm rot}$ measurements.
        }
    \label{fig:Prot_EW_scatter}

\end{figure}

10/15 of our M dwarf sample have tens to hundreds of archival \cak~observations, and we derived our equivalent width measurements from all the spectra to avoid biases from stellar activity on all timescales (i.e., flares, rotation, stellar magnetic activity cycle). Equivalent width lightcurves are shown in Figure~\ref{fig:time_eqw} for the fifteen stars. Variability is observed, especially over the course of a few days, but no definitive signs of cyclic activity due to stellar rotation or the stellar dynamo are observed, nor are they ruled out. Approximately 20\% spreads in the \cak~equivalent widths are typically observed over the course of a few days (see Figures~\ref{fig:ca_profile_overplot} and \ref{fig:Prot_EW_scatter}) and are likely due to flares and rotation of starspots into and out of view. Figure~\ref{fig:ca_profile_overplot} shows the evolution of the \cak~emission profile for GJ 876 over the course of four days and the corresponding change in the corrected \cak~equivalent widths. 

For each target, we computed a mean \cak~equivalent width with an uncertainty equal to the standard deviation of the measurements (Table~\ref{table:targets}). GJ 1132 has only one equivalent width measurement, so the photometric error bar was reported for that measurement. The averages from stars with few observations could be strongly skewed if they coincide with an unresolved episode of high activity. The targets with the fewest observations are GJ 832, GJ 1214, GJ 1132, Proxima Cen, and GJ 213 (see Section~\ref{sec:ObservationsReductions} and Table~\ref{table:targets} for the list of sources for our optical spectra).

Figure~\ref{fig:Prot_EW_scatter} shows the demographics of our sample (stellar rotation period $P_{\rm rot}$ and effective temperature $T_{\rm eff}$) compared to the corrected \cak~equivalent widths and the observed scatter in those values. In general, the fast rotators ($P_{\rm rot}$~$\textless$~7 days) have the largest equivalent widths. Unsurprisingly, these stars are the well-known flare stars AD Leo, EV Lac, AU Mic, and YZ CMi, and they are also the youngest stars ($\textless$ 1 Gyr) in the sample. Between $P_{\rm rot}$~$\sim$~40--85 days, most stars exhibit small equivalent widths, but Proxima Cen (star 10) and GJ 176 (star 4) are outliers. \cite{West2015} find that late M dwarfs (M5--M8) rotating faster than 86 days and early M dwarfs (M0--M4) rotating faster than 26 days exhibit greater optical activity (i.e. an H$\alpha$~emission spectrum and greater \ca~equivalent width; \citealt{Walkowicz2009}). Proxima Cen meets this optical activity criterion, but GJ 176 does not. However, even the ``optically-inactive" stars (H$\alpha$~absorption spectrum) in our sample exhibit UV activity.

The right panel of Figure~\ref{fig:Prot_EW_scatter} shows the scatter observed in the \cak~$W_{\rm \lambda,corr}$ lightcurves normalized to $W_{\rm \lambda,corr}$. The scatter ranges from $\sim$10\%--25\% (although Proxima Cen's scatter is 43\%) with no clear dependence on $T_{\rm eff}$ and a possible positive correlation with $P_{\rm rot}$, although our sample is sparse for $T_{\rm eff}$~$\textless$~3200 K and $P_{\rm rot}$~$\textgreater$~60 days.

\subsection{UV--optical correlations} \label{sec:correlations}

For our 15 M dwarf sample, we compare the average \cak~corrected equivalent widths ($W_{\rm \lambda, corr}$) reported in Table~\ref{table:targets} with UV emission line surface fluxes. By using the corrected \cak~equivalent width (Section~\ref{sec:eqw}) and UV surface fluxes, we attempt to minimize spectral type dependences in our results. We note, however, that the uncertainties in the stellar radii (used to compute the stellar surface flux) are large, and there may be a systematic bias for the lowest mass stars. Direct interferometric measurements yield vastly different radii for mid-M dwarfs of similar effective temperatures; in particular, directly-measured radii are tens of percent larger than radii determined by indirect means \citep{VonBraun2014}. The nine UV lines we used for comparison are listed with their formation temperatures in Table~\ref{table:emlines}. 

\begin{figure*}
   \begin{center}
   
     \subfigure{
          \includegraphics[width=\textwidth]{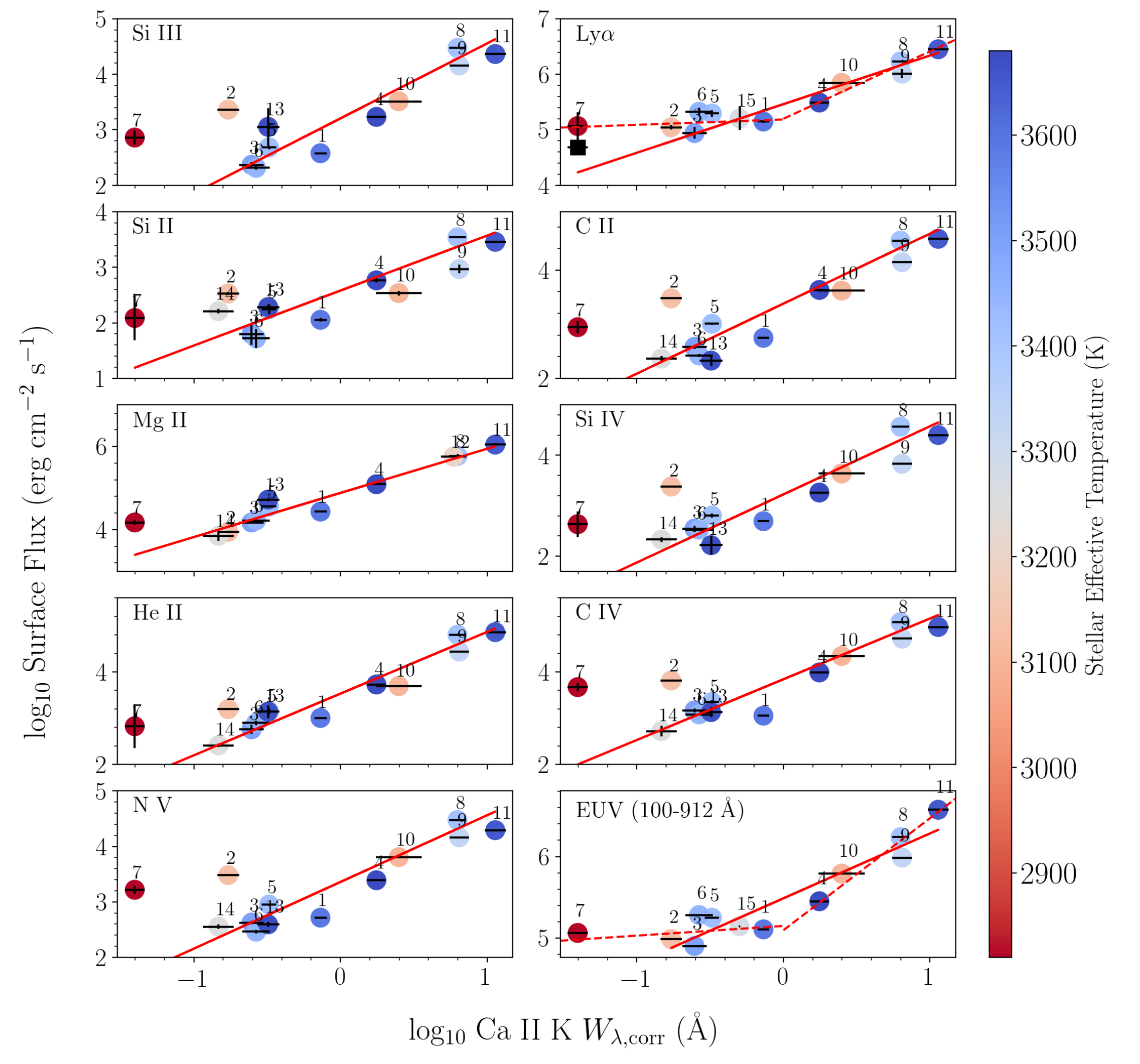}
          }

   \end{center}
    \caption{Ca II K (3933 \AA) corrected equivalent widths ($W_{\rm \lambda,corr}$) and surface fluxes for the nine UV emission lines and total extreme-UV flux. Each point represents a different star (numbered to match Table~\ref{table:targets}) color-coded by effective temperature. In each panel, the solid red line shows a power law fit to the data. In the \Lya~and EUV panels, the dashed red lines show fits applied separately to the ``inactive" and ``active" M dwarfs. The black square in the top right panel shows the uncorrected \Lya~surface flux for GJ 1214 (see \citealt{Youngblood2016}). GJ 1214 was not included in the \Lya--\cak~or EUV--\cak~fits, but was included in all others. The fitted power law parameters are printed in Table~\ref{table:emlines}.
        }
    \label{fig:cak_all_ions}

\end{figure*}

Figure~\ref{fig:cak_all_ions} shows the correlations between $W_{\rm \lambda, corr}$ and surface flux ($F_{\rm S, UV}$) for nine UV emission lines and the extreme-UV flux, which is derived from \Lya~(see Section~\ref{sec:euv_from_cak}), with their power law fits of the form log$_{10}$ $F_{\rm S, UV}$= ($\alpha$ $\times$ log$_{10}$ $W_{\rm \lambda,corr}$) + $\beta$ (Table \ref{table:emlines}). We find statistically significant correlations between \cak~and all nine UV emission lines, which have formation temperatures ranging from 30,000 K to 160,000 K. As expected, the correlations with the least scatter are for \MgII~and \Lya~(0.27 and 0.18 dex, respectively), both optically thick lines with formation temperatures similar to \cak. \ca~has been previously found to correlate positively with \Lya~\citep{Linsky2013} and \MgII~\citep{Walkowicz2009}. Note that due to the large uncertainty in GJ 1214's \Lya~flux \citep{Youngblood2016}, we do not include it in the fit, although its effect on the fit is small. GJ 1214 is an outlier from the fits for all nine UV lines, and GJ 876 is a significant outlier for all but \MgII~and \Lya. Of the seven MUSCLES M dwarfs, GJ 876 exhibited the most frequent and largest flares. These flares were observed in all lines except \MgII~and \Lya, which were observed on different \textit{HST} visits from the other far-UV emission lines, so GJ 876's apparently elevated UV surface fluxes in Figure~\ref{fig:cak_all_ions} may be due to these flares. There is no clear dependence on fit quality (as measured by the Pearson correlation coefficients or the standard deviation of the fit) with the formation temperature of the UV line.

By eye, it appears that some of the \ca--UV correlations in Figure~\ref{fig:cak_all_ions} may be better fit by two power laws with a break around $W_{\rm \lambda, corr}$ $\approx$~1, the transition in this dataset between the ``inactive" ($W_{\rm \lambda, corr}$~$\lesssim$~1) and ``active" ($W_{\rm \lambda, corr}$~$\gtrsim$~1) M dwarfs. In the \Lya~and extreme-UV panels in Figure~\ref{fig:cak_all_ions}, we show two power law fits in addition to the single power law fit. As with the single power law fits, GJ 1214 is excluded. The broken power law fits indicate that for these two UV surface fluxes, the \ca--UV correlations become approximately constant at low stellar activity (Table~\ref{table:emlines}). Greater study of the low-activity M dwarfs is necessary to determine if the UV fluxes become approximately constant in that regime, as the apparent flattening of many of the correlations is largely due to a single star: GJ 1214.

\clearpage

\begin{deluxetable}{cccccccc}
\tabletypesize{\scriptsize}
\tablecolumns{8}
\tablewidth{0pt}
\tablecaption{ Fit parameters for UV surface flux and \cak~equivalent width relations \label{table:emlines}} 
\tablehead{\colhead{Transition name} & 
                  \colhead{Wavelength (\AA)} & 
                  \colhead{log $T_{\rm formation}^a$} & 
                  \colhead{$\alpha$} &
                  \colhead{$\beta$} &
                  \colhead{$\rho$} & 
                  \colhead{$n$} &
                  \colhead{$\sigma$}
                  }
\startdata
Si III & 1206.50 & 4.7 & 1.35$\pm$0.26 & 3.21$\pm$0.16 & 0.80 & 2.0$\times$10$^{-3}$ & 0.59\\
H I~\Lya & 1215.67 & 4.5 (line core) & 0.88$\pm$0.10 & 5.46$\pm$0.06 & 0.95 & 1.1$\times$10$^{-5}$ & 0.18\\
& & ``active" & 1.21$\pm$0.16 & 5.20$\pm$0.12 &  &  & \\
& & ``inactive" & 0.10$\pm$0.17 & 5.19$\pm$0.08 &  &  & \\
Si II & 1260.42, 1264.74, 1265.00 & 4.5 & 0.99$\pm$0.18 & 2.58$\pm$0.11 & 0.82 & 6.7$\times$10$^{-4}$ & 0.42\\
C II$^b$ & 1335.71 & 4.5 & 1.29$\pm$0.23 & 3.38$\pm$0.14 & 0.82 & 6.0$\times$10$^{-4}$ & 0.54\\
Mg II$^c$ &  2796.35,  2803.53 & 4.5 (line core) & 1.06$\pm$0.13 & 4.88$\pm$0.08 & 0.94 & 7.1$\times$10$^{-6}$ & 0.27\\
Si IV & 1393.76, 1402.77 & 4.9 & 1.35$\pm$0.24 & 3.22$\pm$0.14 & 0.84 & 3.6$\times$10$^{-4}$ & 0.54\\
He II & 1640.4$^d$ & 4.9 & 1.33$\pm$0.15 & 3.53$\pm$0.09 & 0.91 & 1.2$\times$10$^{-5}$ & 0.41\\
C IV & 1548.19, 1550.78 & 5.0 & 1.31$\pm$0.26 & 3.84$\pm$0.16 & 0.81 & 8.8$\times$10$^{-4}$ & 0.56 \\
N V & 1238.82, 1242.8060 & 5.2 & 1.2$\pm$0.26 & 3.36$\pm$0.15 & 0.77 & 1.9$\times$10$^{-3}$ & 0.55\\
EUV$^e$ & 100--912 & -- & 0.80$\pm$0.09 & 5.49$\pm$0.06 & 0.94 & 1.5$\times$10$^{-5}$ & 0.18\\
& & ``active" & 1.37$\pm$0.18 & 5.10$\pm$0.14 &  &  & \\
& & ``inactive" & 0.12$\pm$0.18 & 5.15$\pm$0.08 &  &  & \\
\enddata
\tablecomments{All relations have the form log$_{10}$ $F_{\rm S, UV}$= ($\alpha$ $\times$ log$_{10}$ $W_{\rm \lambda,corr}$) + $\beta$, where $F_{\rm S, UV}$ is the surface flux of the UV emission line in erg cm$^{-2}$ s$^{-1}$ and $W_{\rm \lambda,corr}$ is the Ca II K equivalent width in \AA. $\rho$~is the Pearson correlation coefficient, $n$ is the probability of no correlation, and $\sigma$~is the standard deviation of the data points about the best-fit line (dex). The additional fit parameters for \Lya~and EUV surface flux apply separately to the ``active" ($W_{\rm \lambda,corr}$~$\textgreater$~1) and ``inactive" M dwarfs ($W_{\rm \lambda,corr}$~$\textless$~1). Combining the active and inactive fit components, $\sigma$~= 0.12 for \Lya, and $\sigma$~= 0.13 for the EUV. }
\tablenotetext{a}{Formation temperatures are from the CHIANTI database \citep{Dere1997,Landi2013}.}
\tablenotetext{b}{Does not include the 1334.54 \AA~line due to significant ISM absorption.}
\tablenotetext{c}{Fluxes corrected for 30\% ISM absorption (see Section~\ref{sec:UVopticaldata} and \citealt{Youngblood2016}).}
\tablenotetext{d}{Average wavelength of the multiplet.}
\tablenotetext{e}{Extreme-UV (EUV) fluxes calculated from \Lya~fluxes using scaling relations from \cite{Linsky2014}.}
\end{deluxetable}

\clearpage

\begin{deluxetable}{ccc}
\tablecolumns{3}
\tablewidth{0pt}
\tablecaption{ Formulae for estimating extreme-UV fluxes from \Lya~and \cak \label{table:CaK_EUV}} 
\tablehead{\colhead{Flux in Wavelength} &
                   \colhead{} &
                   \colhead{}  \\
                  \colhead{Band at 1 AU} &
                  \colhead{\cite{Linsky2014}} &
                  \colhead{} \\
                  \colhead{(erg cm$^{-2}$ s$^{-1}$)} &
                  \colhead{Table 5} & 
                  \colhead{This work$^b$}
                  }
\startdata
$F$(100--200 \AA) & 10$^{-0.491}$~$\cdot$~$F$(\Lya) & 10$^{4.97\pm0.06}$~$\cdot$~$W_{\rm \lambda,corr}^{0.88\pm0.10}$~$\cdot$~(4.65$\times$10$^{-3}$$R_{\star}$)$^{2.0}$ \\ [7pt]
$F$(200--300 \AA) & 10$^{-0.548}$~$\cdot$~$F$(\Lya) & 10$^{4.91\pm0.06}$~$\cdot$~$W_{\rm \lambda,corr}^{0.88\pm0.10}$~$\cdot$~(4.65$\times$10$^{-3}$$R_{\star}$)$^{2.0}$ \\[7pt]
$F$(300--400 \AA) & 10$^{-0.602}$~$\cdot$~$F$(\Lya) & 10$^{4.86\pm0.06}$~$\cdot$~$W_{\rm \lambda,corr}^{0.88\pm0.10}$~$\cdot$~(4.65$\times$10$^{-3}$$R_{\star}$)$^{2.0}$ \\[7pt]
$F$(400--500 \AA) & 10$^{-2.294}$~$\cdot$~$F$(\Lya)$^{1.258}$ & 10$^{4.57\pm0.08}$~$\cdot$~$W_{\rm \lambda,corr}^{1.11\pm0.13}$~$\cdot$~(4.65$\times$10$^{-3}$$R_{\star}$)$^{2.52}$ \\[7pt]
$F$(500--600 \AA) & 10$^{-2.098}$~$\cdot$~$F$(\Lya)$^{1.572}$ & 10$^{6.49\pm0.09}$~$\cdot$~$W_{\rm \lambda,corr}^{1.38\pm0.16}$~$\cdot$~(4.65$\times$10$^{-3}$$R_{\star}$)$^{3.14}$ \\[7pt]
$F$(600--700 \AA) & 10$^{-1.920}$~$\cdot$~$F$(\Lya)$^{1.240}$ & 10$^{4.85\pm0.07}$~$\cdot$~$W_{\rm \lambda,corr}^{1.09\pm0.12}$~$\cdot$~(4.65$\times$10$^{-3}$$R_{\star}$)$^{2.48}$ \\[7pt]
$F$(700--800 \AA) & 10$^{-1.894}$~$\cdot$~$F$(\Lya)$^{1.518}$ & 10$^{6.39\pm0.09}$~$\cdot$~$W_{\rm \lambda,corr}^{1.34\pm0.15}$~$\cdot$~(4.65$\times$10$^{-3}$$R_{\star}$)$^{3.04}$ \\[7pt]
$F$(800--912 \AA) & 10$^{-1.811}$~$\cdot$~$F$(\Lya)$^{1.764}$ & 10$^{7.82\pm0.11}$~$\cdot$~$W_{\rm \lambda,corr}^{1.55\pm0.18}$~$\cdot$~(4.65$\times$10$^{-3}$$R_{\star}$)$^{3.53}$ \\[7pt]
\enddata
\tablecomments{$F$(\Lya) is the \Lya~flux at 1 AU (erg cm$^{-2}$ s$^{-1}$) and $W_{\rm \lambda,corr}$ is the \cak~corrected equivalent width (Equation~\ref{eq:corrfactor}).}
\tablenotetext{b}{Substituted \Lya--$W_{\rm \lambda,corr}$ scaling relation from Table~\ref{table:emlines} into relation from \cite{Linsky2014}; $R_{\star}$ in units of $R_{\odot}$; uncertainties propagated from fit uncertainties in Table~\ref{table:emlines}.}
\end{deluxetable}

\subsubsection{Estimating the Extreme-UV Spectrum from \cak} \label{sec:euv_from_cak}

Predicting an M dwarf's \Lya~flux is important not only because \Lya~constitutes a major fraction of the far-UV flux, but it is also a means for estimating the extreme-UV spectrum, which currently cannot be observed for any star except the Sun. In Figure~\ref{fig:cak_all_ions} and Table~\ref{table:emlines}, we have determined the scaling relation between the total extreme-UV flux (100--912 \AA) and $W_{\rm \lambda,corr}$ for our stars. The fit is very similar to the \Lya--\cak~best-fit line, because the extreme-UV fluxes were derived from scaling relations with \Lya~from \cite{Linsky2014}.

In Table~\ref{table:CaK_EUV}, we substitute our \ca--\Lya~scaling relation into the \Lya--EUV scaling relations in Table 5 of \cite{Linsky2014} to allow the reader to directly reconstruct the extreme-UV spectrum from a \cak~$W_{\rm \lambda,corr}$ measurement. We have also reprinted the scaling relations from \cite{Linsky2014} in Table~\ref{table:CaK_EUV}, although they have been simplified for brevity. 

\subsubsection{Estimating the Uncertainties on Derived UV Surface Fluxes} \label{sec:error in UV calcs}

Here we estimate the uncertainties in the UV fluxes estimated from \cak~using the presented scaling relations (Table~\ref{table:emlines} and \ref{table:CaK_EUV}). Important sources of error include the non-simultaneity of our UV and optical observations, the uncertainties in the \cak~and UV~measurements, the uncertainties in the stellar radii and distances (although the uncertainties in distances for these nearby stars are small) that are used to calculate surface flux, and the uncertainties in the stellar effective temperatures, surface gravities, and metallicities used to calculate the equivalent width correction factors. The error bars on the $W_{\rm \lambda,corr}$ values come from the dispersion in the many equivalent width measurements made for each target (with the exception of GJ 1132, which has only one measurement), and these error bars range from $\sim$10\%--25\%, although Proxima Cen's $W_{\rm \lambda,corr}$ error bar is close to 45\%.

Of the nine UV emission lines, \Lya~and \MgII~have the least scatter about the best-fit line (0.18 and 0.27 dex, respectively). This is important, because these two emission lines comprise the majority of the far-UV and near-UV emission line flux, respectively, for the M dwarfs. Considering $F$(emission line)/$\sum F$(nine emission lines) for the nine stars with measurements for all nine emission lines, $F$(\Lya)/$\sum F$(nine emission lines) = 65\%--91\% and $F$(\MgII)/$\sum F$(nine emission lines) = 6\%--27\%. The other emission lines comprise smaller percentages of the total emission line flux: $\sim$0.1\%--5\% each. For example, the scatter in the \SiIII--\cak~scaling relation is the largest ($\sigma$ = 0.59 dex), but $F$(\SiIII)/$\sum F$(nine emission lines) = 0.09\%--1.7\%.

Due to the limited number of extreme-UV observations of M dwarfs, quantifying the true uncertainty in the calculated extreme-UV spectrum is challenging. Here we estimate the uncertainty and list all the main sources of error. \cite{Linsky2014} used \textit{Extreme UltraViolet Explorer} (\textit{EUVE}) observations (100--400 \AA) of six M dwarfs with \textit{HST}/STIS \Lya~observations, including AU Mic, Proxima Cen, AD Leo, EV Lac, and YZ CMi. The 400--912 \AA~spectra were provided by semi-empirical models \citep{Fontenla2014}. Scaling relations for eight $\sim$100 \AA~bandpasses in the extreme-UV were derived, with dispersions of 13--24\%. \cite{Linsky2014} describe three sources of uncertainty in their technique: errors in the extreme-UV fluxes, errors in the reconstructed \Lya~fluxes, and errors associated with stellar variability (the extreme-UV and \Lya~observations were not simultaneous). The observed dispersion (13--24\%) in the scaling relations is surprisingly small given the expected magnitudes of the three uncertainty sources. \cite{Linsky2014} attribute this to the avoidance of \textit{EUVE} observations containing flares.

The scatter in our own \cak--\Lya~relationship is surprisingly small ($\sigma$ = 0.18 dex) given the uncertainties listed in the first paragraph of this subsection. The $W_{\rm \lambda,corr}$ error bars range from $\sim$10\%--30\%, and \cite{Youngblood2016} find that the uncertainties in the reconstructed \Lya~fluxes range from $\sim$5\%--20\% for moderate-to-high S/N observations. GJ 1214 was the lowest S/N observation and had a $\sim$100\% uncertainty in the reconstructed flux, but was not included in the \cak--\Lya~fit.

Starting with a $W_{\rm \lambda,corr}$ measurement with an assumed 30\% uncertainty and using the \cak--\Lya~scaling relation from Table~\ref{table:emlines}, we find that the propagated uncertainty in the calculated \Lya~flux is unchanged, indicating that the uncertainty in the \cak~equivalent width dominates. Using the calculated \Lya~flux to estimate the extreme-UV spectrum and adding the 30\% uncertainty in quadrature with the 24\% dispersion from \cite{Linsky2014} yields an uncertainty of $\sim$40\% for the resulting extreme-UV flux. 

We have assumed no additional uncertainty due to the variations in metallicity of our target stars. Metallicity variations likely have the largest impact on the \HI~\Lya--\ca~relations, but it appears that this effect is negligible compared to other sources of scatter for our near-solar metallicity (-0.5 $\leq$~[Fe/H]~$\leq$~0.5) target stars. The metallicity effect could become significant for metal-poor stars ([Fe/H] $\textless$~-1), where the relative abundance of Ca with respect to H is approximately an order of magnitude less than for solar-metallicity stars, and we caution against applying these correlations to stars with any physical parameters beyond the bounds of our 15 star sample. The addition of metal-poor M dwarfs into the sample would allow for a determination of the effect of metallicity on these UV--optical correlations.

\section{Energetic proton estimation from UV flares} \label{sec:uvproton}
\textit{HST} has observed dozens of spectrally and temporally resolved far-UV flares from M dwarfs (\citealt{Loyd2014,France2016,Loyd2017inprep}, in preparation), which can be used to constrain the time-dependent energy input into the upper atmospheres of orbiting exoplanets. Energetic particles from stellar eruptive events are not frequently included in these energy budgets, because there are no observational constraints for stars other than the Sun. Existing solar correlations between SXRs and protons detected near Earth (i.e. \citealt{Belov2007,Cliver2012a}) cannot be directly applied to \textit{HST}'s UV flares, because we do not know the energy partition between stellar UV emission lines and SXRs during flares. Thus we have developed a new scaling relation between energetic protons detected near Earth and UV flares from the Sun.

Ideally, we would determine the relationship between energetic protons detected by the \textit{GOES} satellites and far-UV spectra of the Sun, because this would be directly comparable to the flares detected by \textit{HST}. However, there are no disk-integrated, high-cadence solar observations of UV emission lines within \textit{HST}'s STIS or COS nominal far-UV spectral ranges (1150--1700 \AA). We elected to use an extreme-UV emission line from high-cadence solar irradiance measurements as a proxy for far-UV emission lines. \textit{SDO}/EVE measured the solar spectral irradiance of the Sun from 50\,--\,370 \AA~in the MEGS-A channel at 10 second cadence from 2010--2014. The only high-S/N ion with a formation temperature similar to those accessible by \textit{HST} STIS/COS in the far-UV is \HeII~at 304 \AA~(log $T_{\rm formation}$ = 4.9; CHIANTI; \citealt{Dere1997,Landi2013}). The far-UV flare tracers \SiIV~(1393, 1402 \AA; log $T_{\rm formation}$ = 4.9) and \HeII~(1640 \AA; log $T_{\rm formation}$ = 4.9) observed by \textit{HST} have similar formation temperatures to \HeIIEUV. After determining the relationship between $\textgreater$10 MeV proton flux and \HeIIEUV, we use the semi-empirical model of GJ 832 (M1.5 V; \citealt{Fontenla2016}) to scale \HeIIEUV~flux to \SiIV~and \HeIIFUV~flux. In Section~\ref{sec:app_to_flare}, we estimate the proton flux from a \SiIV~flare observed with \textit{HST} from the M4 dwarf GJ 876, and in Section~\ref{subsubsec:limitations}, we discuss the limitations of applying these proton scaling relationships to flares on M dwarfs.

\begin{figure}[t]
   \begin{center}
   
     \subfigure{
          \includegraphics[width=\textwidth]{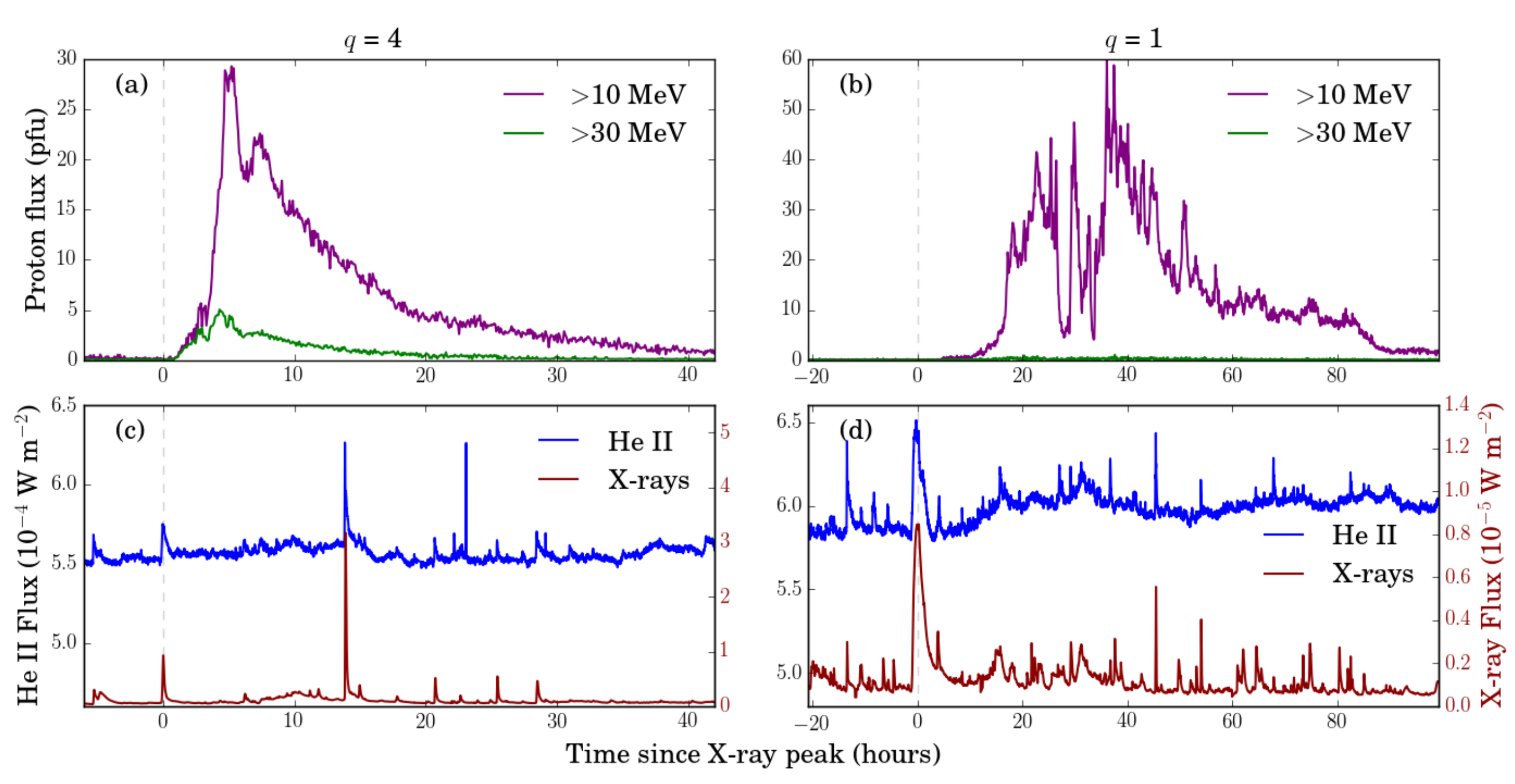}
          }
               
   \end{center}
    \caption{Examples of SXR and \HeII~(304 \AA) flares and corresponding proton enhancements. Panels (a) and (c) show an example where we have high confidence ($q$ = 4) in the \HeIIEUV~flux measurement and that the events are associated. This SXR flare peaked on 2013 December 12 18:02:35 UTC. Panels (b) and (d) show an example event that we have low confidence ($q$ = 1) in both the measured \HeIIEUV~flux and in the association between the flare and proton enhancement. This SXR flare peaked on 2012 August 31 20:44:05 UTC.  
        }
    \label{fig:example_flares}

\end{figure}  

\subsection{The solar flare and proton enhancement sample} \label{sec:solar_sample}

We identified 36 proton enhancements in the \textit{GOES} $\textgreater$10 MeV and $\textgreater$30 MeV proton channels with an associated SXR (1--8 \AA~with \textit{GOES}) and \HeIIEUV~flare. Our confidence levels for the proton--flare association varied, and we assigned each event a quality index ($q$) ranging from 1 (lowest confidence) to 4 (highest confidence). The number of events assigned to $q$ = 1, 2, 3, and 4 is 7, 11, 7, and 11, respectively. Reasons for a low $q$ include difficulty in identifying a proton enhancement's precursor flare and/or reliably measuring the properties of both fluxes. There can be too many candidate precursor flares, low S/N, difficulty in defining the beginning and end of the proton and flare events, and uncertainty in measuring background flux level. Figure~\ref{fig:example_flares} shows an example high confidence event ($q$ = 4) and a low confidence event ($q$ = 1). For the $q$ = 1 example event, the onset of the $\textgreater$10 MeV protons after the photon flares is much more gradual than most events, and the $\textgreater$30 MeV proton flux shows no significant rise. The \HeIIEUV~lightcurve also shows many events, which may be confused, and estimation of the background level is challenging. 

\begin{figure}[t]
   \begin{center}
   
     \subfigure{
          \includegraphics[width=\textwidth]{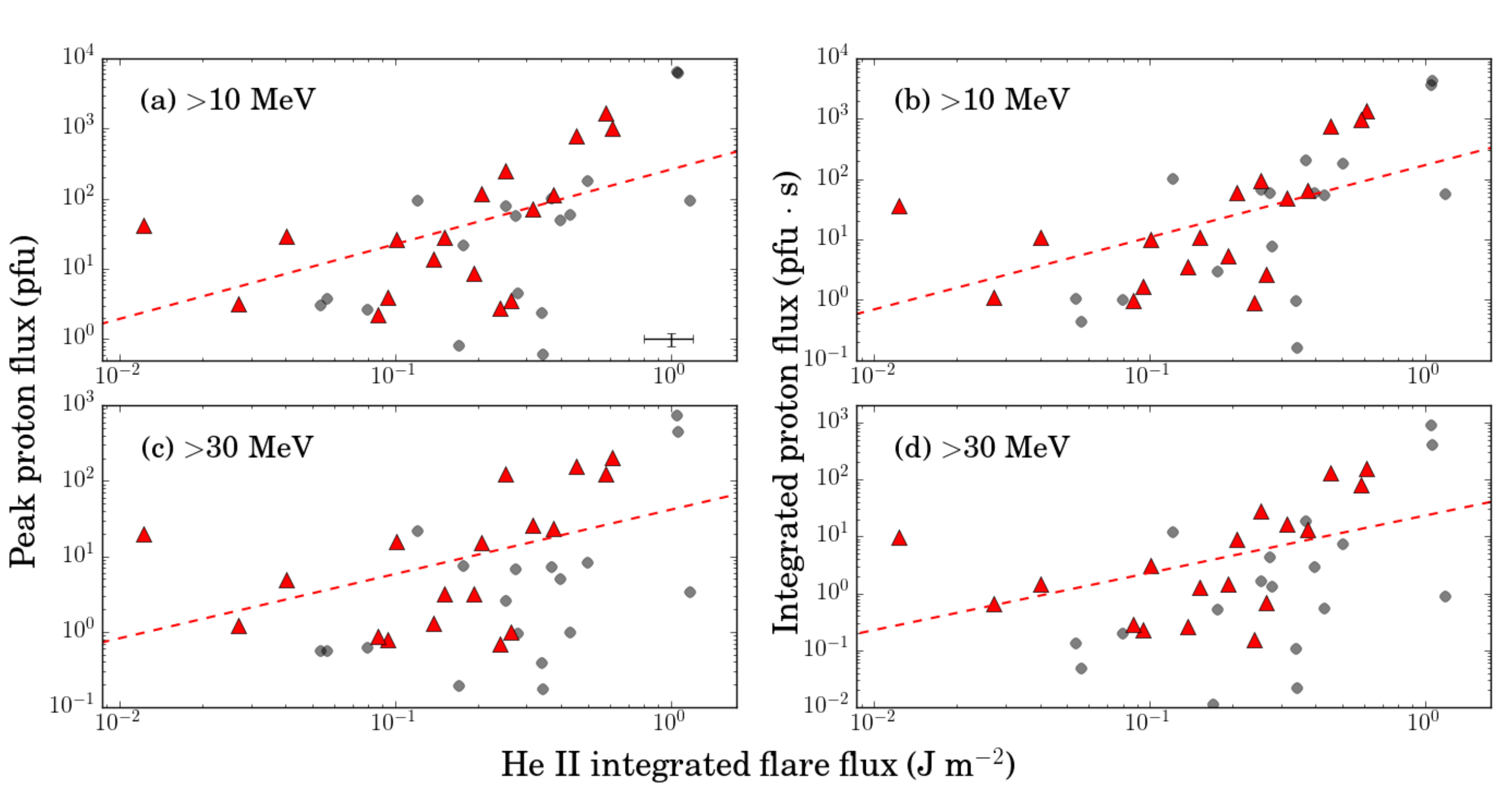}
          }
          
   \end{center}
    \caption{Relations between background-subtracted \HeII~(304 \AA) flare fluence and peak flux and fluence of likely-associated proton enhancements. The red triangles represent data points with a quality factor $q$~$\geq$~3 and the dashed red lines show power law fits to those points. The gray circles represent the data points $q$ $\leq$~2 and were excluded from the fits. Panels (a) and (c) show the peak proton flux (background subtracted), and (b) and (d) show the proton fluence (background subtracted). 1 pfu = 1 proton cm$^{-2}$ s$^{-1}$ sr$^{-1}$. The fit parameters are listed in Table~\ref{table:prot_correlations}, and representative 20\% error bars for the $q$~$\textgreater$~3 data points are shown in the lower right corner of panel (a).
        }
    \label{fig:HeIItotfluxbacksub_prottotfluxbacksub}

\end{figure}  

We find that the proton, UV, and SXR flux errors are dominated by systematics in defining a background flux level to subtract and the duration of the flare. We applied a linear fit to the background around each event. For some events, this was straightforward, but for others, this method likely increases uncertainty in the measurement. We determined the beginning and end of the flare by where the lightcurve intersected the background level fit. To estimate the uncertainties in our measurements, we re-measured four of the 36 events three times, each time slightly changing the background fit and the beginning and end of the flare. We estimate the uncertainty in the background-subtracted flux to range from 10\%--300\% for SXR, 10\%--200\% for \HeIIEUV, and 10\%--30\% for the protons. These ranges are a reflection of the varying quality factors: $q$~$\textless$~3 events have the largest uncertainties and $q$~$\geq$~3 events have the smallest uncertainties. If we consider just the fluence (no background subtraction) or the peak fluxes, the uncertainties drop to 10\% for the SXRs and \HeIIEUV, indicating background subtraction and not flare duration is the dominant uncertainty.

The 36 \HeIIEUV~flares have mean durations 5.3 $\pm$~4.6 hours and on average peak a few minutes after the SXR flare peak, although there is large scatter in this average. When examining only the $q$~$\geq$~3 events, the average \HeIIEUV~peak occurs $\sim$5 minutes before the SXR peak. \cite{Kennedy2013} used \HeIIEUV~to trace the impulsive phase flares, and found \HeIIEUV~to peak 1--4 minutes before the SXR peak, which traces the more gradual phase of the flare, and \cite{Milligan2012} found for an X-class flare that \HeIIEUV~peaked 18 s before the \textit{GOES} SXRs. The associated proton enhancements begin within about 2 hours of the SXR peak, and on average last for 3 days. Part of the scatter in the \HeIIEUV--proton relationships presented in Figure~\ref{fig:HeIItotfluxbacksub_prottotfluxbacksub} is due to overlapping flares with contributing protons. In the duration of a proton enhancement, the same or another active region could flare again and accelerate protons that reach Earth. Many of the $\textgreater$10 MeV enhancements have many peaks over the course of several days, while the $\textgreater$30 MeV enhancements typically only have a rapid initial peak and a gradual decline. The peak proton fluxes between the two channels generally do not coincide temporally; the $\textgreater$10 MeV peak particle flux typically occurs after the $\textgreater$30 MeV peak. The energy-dependent arrival times of protons are not completely explained by differing speeds; it is thought that higher-energy protons are accelerated with electrons close to the solar surface, and that lower-energy protons are either accelerated at a later time (farther from the solar surface) or their escape from the Sun is delayed due to trapping by shocks \citep{Krucker2000,Xie2016}.

\begin{figure}[t]
   \begin{center}
   
     \subfigure{
          \includegraphics[width=0.5\textwidth]{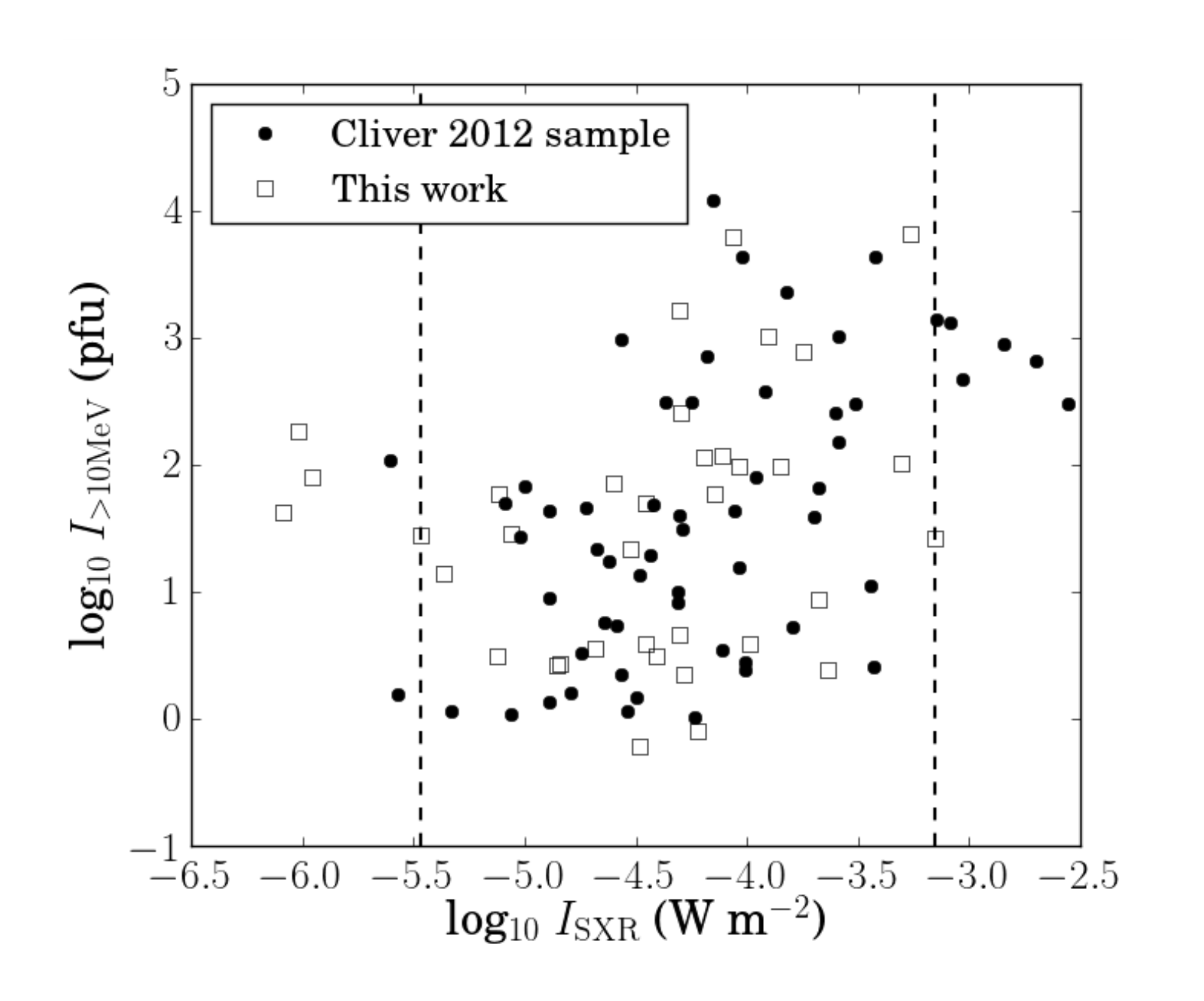}
          }
               
   \end{center}
    \caption{SXR (1-8 \AA) peak flare intensity (not background subtracted per \textit{GOES} standards; $I_{\rm SXR}$; W m$^{-2}$) and peak $\textgreater$10 MeV proton flux ($I_{\rm \textgreater 10MeV}$; pfu). Solid black circles show the 58 events presented in Figure 2 of \cite{Cliver2012a}, and the unfilled black squares show the 36 events presented in this work. The two vertical dashed lines enclose the data points that cover the same SXR parameter range.
        }
    \label{fig:Cliver_comparison}

\end{figure}  

\begin{figure}[t]
   \begin{center}

     \subfigure{
          \includegraphics[width=0.5\textwidth]{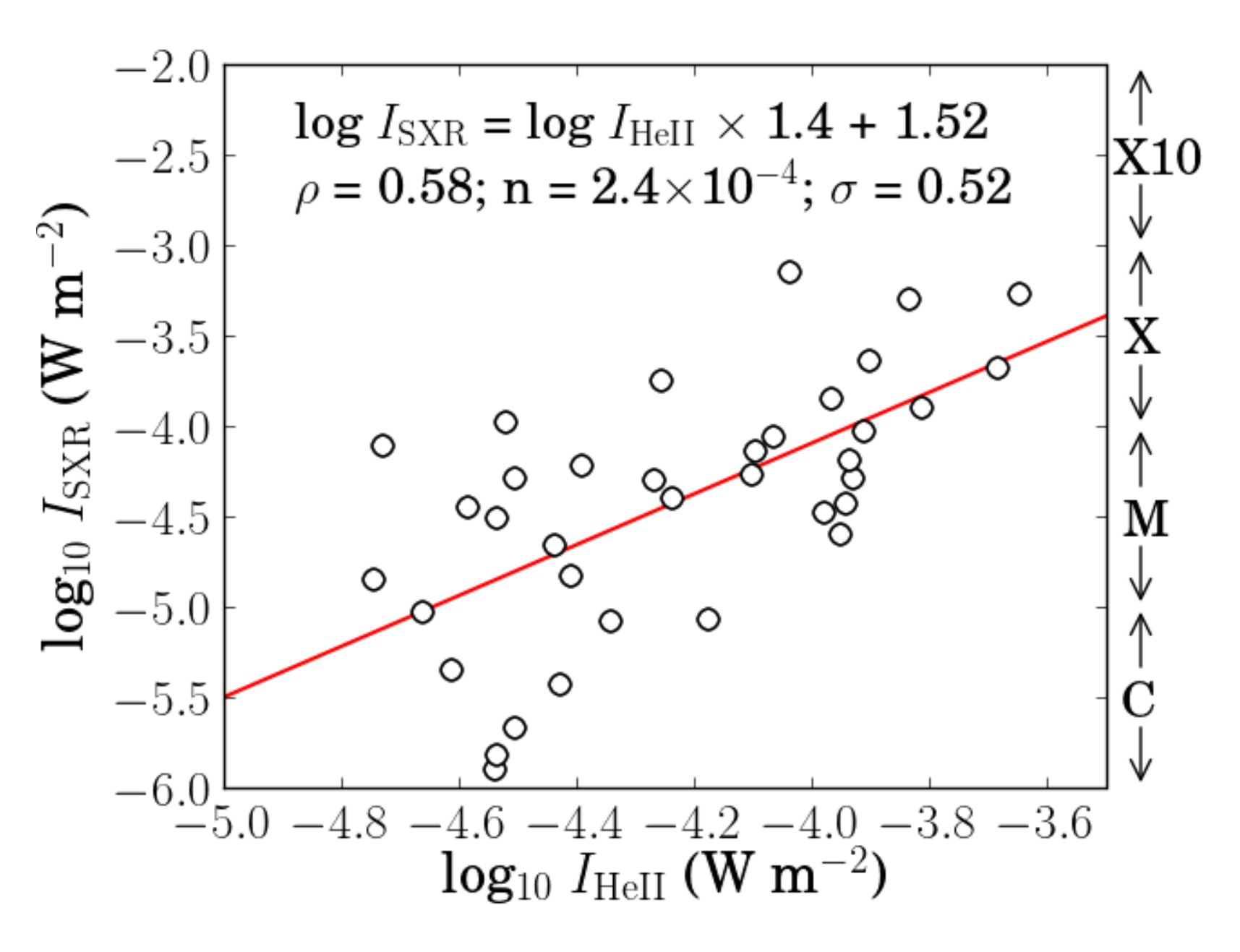}
          }
          
   \end{center}
    \caption{\HeIIEUV~peak flare intensity (background subtracted; $I_{\rm HeII}$) and SXR (1-8 \AA) peak flare intensity (not background subtracted per \textit{GOES} standards; $I_{\rm SXR}$) of the 36 events (white circles). The right side of the plot is marked with the \textit{GOES} SXR flare classification scheme, which is also described in Table~\ref{table:GOES classification}.
        }
    \label{fig:Xray_HeII_peak}

\end{figure}  

In their sample of 58 SXR and proton enhancement events, \cite{Cliver2012a} found a statistically significant correlation ($\rho$~= 0.52, $n$ = 3$\times$10$^{-5}$), but our 36 event sample yields $\rho$~= 0.18, $n$ = 0.3 (Figure~\ref{fig:Cliver_comparison}). If we force both samples to cover the same parameter space (enclosed in the vertical lines in Figure~\ref{fig:Cliver_comparison}), the correlation coefficients become similar: $\rho$~= 0.41, $n$ = 0.003 for 50 of \cite{Cliver2012a} events, and $\rho$~= 0.35, $n$ = 0.05 for 33 of our events. The different $n$ values are due to the different sample sizes.

\subsection{Solar UV--proton scaling relations} \label{sec:solar_scalingrelations}

We observe statistically significant correlations between \HeIIEUV~fluence (time-integrated flux with units J m$^{-2}$) and proton fluence and peak proton flux (all fluence and flux values background subtracted; Figure~\ref{fig:HeIItotfluxbacksub_prottotfluxbacksub}). The data were fitted with power laws (log$_{10}$ $x$ = $\alpha$~$\times$ log$_{10}$ $y$ + $\beta$), and the parameters for the fits are presented in Table~\ref{table:prot_correlations}. The scatter in these fits is large ($\sigma$ = 0.75--0.84 dex).

The near identical formation temperatures of \HeIIEUV, \SiIV, and \HeIIFUV~suggests the ratios of their fluxes are likely to be similar for all stars with metallicities similar to solar values. Thus, we estimated these ratios using synthetic spectra of the Sun (CHIANTI; \citealt{Dere1997,Landi2013}) and GJ 832 \citep{Fontenla2016}. For the Sun, $F_{\rm SiIV}$/$F_{\rm HeII, 304}$ = 0.117, and $F_{\rm HeII, 1640}$/$F_{\rm HeII, 304}$ = 0.0304. 
For GJ 832, $F_{\rm SiIV}$/$F_{\rm HeII, 304}$ = 0.137, and $F_{\rm HeII, 1640}$/$F_{\rm HeII, 304}$ = 0.0291. The ratios for these two stars are very similar, so we will use the GJ 832 ratios in the subsequent analysis. However, we note that the ratios could change by a factor of a few during a flare as estimated by comparing flux ratios of various lines for the active and quiet Sun models listed in Table 1 of \cite{Fontenla2016}. More M dwarf atmosphere models calculated at a range of activity levels would be valuable.

We apply the quiescence ratios to our proton--\HeIIEUV~relations (Table~\ref{table:prot_correlations}) to relate \SiIV~and \HeIIFUV~flare flux to proton enhancements. Using one of the correlations listed in Table~\ref{table:prot_correlations}, and replacing $F_{\rm HeII, 304}$ with $F_{\rm SiIV}$, we find that:

\begin{equation}
{\rm log}~I_{\rm \textgreater10 MeV} = (1.06\pm0.21) \times~{\rm log}~F_{\rm SiIV} + (3.34\pm0.25),
\label{eq:prot_peak_SiIV}
\end{equation}

\noindent where $F_{\rm SiIV}$ is the background-subtracted \SiIV~(1393, 1402 \AA) flare fluence (J m$^{-2}$) as would be observed at 1 AU from the star, and $I_{\rm \textgreater10 MeV}$ is the background-subtracted peak $\textgreater$10 MeV proton enhancement intensity (pfu; 1 pfu = 1 proton cm$^{-2}$ s$^{-1}$ sr$^{-1}$) as would also be observed at 1 AU. If instead the proton fluence (F$_{\textgreater10MeV}$, [pfu $\cdot$~s]) rather than peak proton flux is used, the relationship becomes 

\begin{equation}
{\rm log}~F_{\rm \textgreater10 MeV} = (1.20\pm0.26) \times~{\rm log}~F_{\rm SiIV} + (3.27\pm0.31).
\label{eq:prot_flux_SiIV}
\end{equation}

\noindent Similarly, we can derive $I_{\rm \geq10 MeV}$ and $F_{\rm \geq10 MeV}$ from $F_{\rm HeII, 1640}$:

\begin{equation}
{\rm log}~I_{\rm \textgreater10 MeV} = (1.06\pm0.21) \times~{\rm log}~F_{\rm HeII, 1640} + (4.05\pm0.36),
\label{eq:prot_peak_HeIIFUV}
\end{equation}

\noindent and

\begin{equation}
{\rm log}~F_{\rm \textgreater10 MeV} = (1.20\pm0.26) \times~{\rm log}~F_{\rm HeII, 1640} + (4.07\pm0.45).
\label{eq:prot_flux_HeIIFUV}
\end{equation}

\subsection{UV-based \textit{GOES} flare classification} \label{sec:GOES_class_from_UV}

When using Equations~\ref{eq:prot_peak_SiIV}--\ref{eq:prot_flux_HeIIFUV}, it is important to note that not all far-UV flares will be accompanied by particle events. Less energetic flares are less likely to produce particles. The solar relation was quantified by \cite{Yashiro2006} using the \textit{GOES} SXR classification scheme as the metric of flare strength. Specifically, they associated a probability of a CME with each GOES flare class. To relate this to far-UV data, we used our 36 events to fit an empirical power law between peak SXR intensity and peak \HeIIEUV~intensity (W m$^{-2}$; Figure~\ref{fig:Xray_HeII_peak}):

\begin{equation}
{\rm log}~I_{\rm SXR} = (1.40\pm0.09) \times~{\rm log}~I_{\rm HeII, 304} + (1.52\pm1.62).
\label{eq:Xray_HeII_peak}
\end{equation}

\noindent The Pearson correlation coefficient is 0.58 with $n$ = 2.2 $\times$~10$^{-4}$ for the 36 events used in this analysis. The scatter about the best fit line in Figure~\ref{fig:Xray_HeII_peak} is 0.52 dex, so the resulting classifications for M dwarf flares will be accurate within a factor of a few. This level of accuracy will be problematic for small flares ($\textless$X class), but less so for large flares due to the $\sim$100\% chance of associated proton enhancement (Table~\ref{table:GOES classification}). Using our conversion ratio ($F_{\rm SiIV}$/$F_{\rm HeII, 304}$ = 0.137) and assuming $F_{\rm SiIV}$/$F_{\rm HeII, 304}$ $\sim$ $I_{\rm SiIV}$/$I_{\rm HeII, 304}$, we can estimate the peak SXR flux from the peak \SiIV~and \HeIIFUV~flux:

\begin{equation}
{\rm log}~I_{\rm SXR} = (1.4\pm0.1) \times~{\rm log}~I_{\rm SiIV} + (2.7\pm1.6),
\label{eq:Xray_SiIV_peak}
\end{equation}

\noindent and

\begin{equation}
{\rm log}~I_{\rm SXR} = (1.4\pm0.1) \times~{\rm log}~I_{\rm HeII,1640} + (3.7\pm1.6).
\label{eq:Xray_HeIIFUV_peak}
\end{equation}

\noindent Approximately 20\% of C-class flares ($F_{\rm SXR}$ = 10$^{-6}$--10$^{-5}$ W m$^{-2}$ at 1 AU) and $\sim$100\% of X3 or greater class flares ($F_{\rm SXR}$~$\geq$3 $\times$~10$^{-4}$ W m$^{-2}$ at 1 AU) have associated CMEs (\citealt{Yashiro2006}; Table~\ref{table:GOES classification}). Recall that $F_{\rm SXR}$ is the peak 1-8 \AA~flare intensity measured at 1 AU from the star and is not corrected for pre-flare flux levels. An X-class or greater flare ($\geq$10$^{-4}$ W m$^{-2}$ at 1 AU) corresponds to any \SiIV~flare with a background-subtracted peak intensity value $\geq$1.6 $\times$~10$^{-5}$ W m$^{-2}$ = 1.6$\times$~10$^{-2}$ erg cm$^{-2}$ s$^{-1}$ at 1 AU, and to any \HeIIFUV~flare with a background-subtracted peak intensity value $\geq$3.2 $\times$~10$^{-6}$ W m$^{-2}$ = 3.2 $\times$~10$^{-3}$ erg cm$^{-2}$ s$^{-1}$ at 1 AU.

\begin{deluxetable}{llccccc}
\tablecolumns{7}
\tablewidth{0pt}
\tablecaption{Correlations between UV flare fluence ($F$) and the peak intensity ($I$) and fluence ($F$) of energetic protons during solar flares. \label{table:prot_correlations}} 
\tablehead{\colhead{$x$} & 
                  \colhead{$y$} &
                  \colhead{$\alpha$} & 
                  \colhead{$\beta$} & 
                  \colhead{$\rho$} &
                  \colhead{$n$} & 
                  \colhead{$\sigma$}
                  }
\startdata
$F_{\rm HeII, 304}$ & $I_{\geq 10 \rm MeV}$ & 1.06$\pm$0.21 & 2.42$\pm$0.17 & 0.83 & 2.2$\times$10$^{-5}$ & 0.76\\
$F_{\rm HeII, 304}$ & $I_{\geq 30 \rm MeV}$ & 0.85$\pm$0.20 & 1.62$\pm$0.16 & 0.80 & 6.5$\times$10$^{-5}$ & 0.75 \\
$F_{\rm HeII, 304}$ & $F_{\geq 10 \rm MeV}$ & 1.20$\pm$0.26 & 2.23$\pm$0.21 & 0.84 & 1.0$\times$10$^{-5}$ & 0.84 \\
$F_{\rm HeII, 304}$ & $F_{\geq 30 \rm MeV}$ & 1.01$\pm$0.25 & 1.37$\pm$0.20 & 0.82 & 3.6$\times$10$^{-5}$ & 0.84 \\
\hline
$F_{\rm SiIV}$ & $I_{\geq 10 \rm MeV}$ & 1.06$\pm$0.21 & 3.34$\pm$0.25 \\
$F_{\rm SiIV}$ & $F_{\geq 10 \rm MeV}$ & 1.20$\pm$0.26 & 3.27$\pm$0.31 \\
$F_{\rm HeII,1640}$ & $I_{\geq 10 \rm MeV}$ & 1.06$\pm$0.21 & 4.05$\pm$0.36 \\
$F_{\rm HeII,1640}$ & $F_{\geq 10 \rm MeV}$ & 1.20$\pm$0.26 & 4.07$\pm$0.45 \\
\enddata
\tablecomments{The power law fits (log$_{10}$ $x$ = $\alpha$~$\times$~log$_{10}$ $y$ + $\beta$) are based on the $q$~$\geq$~3 background-subtracted data (Figure~\ref{fig:HeIItotfluxbacksub_prottotfluxbacksub}). $\rho$~is the Pearson correlation coefficient, $n$ is the probability of no correlation, and $\sigma$~is the standard deviation of the data points about the best-fit line (dex).}

\end{deluxetable}

\begin{figure}[t]
   \begin{center}
   
     \subfigure{
          \includegraphics[width=\textwidth]{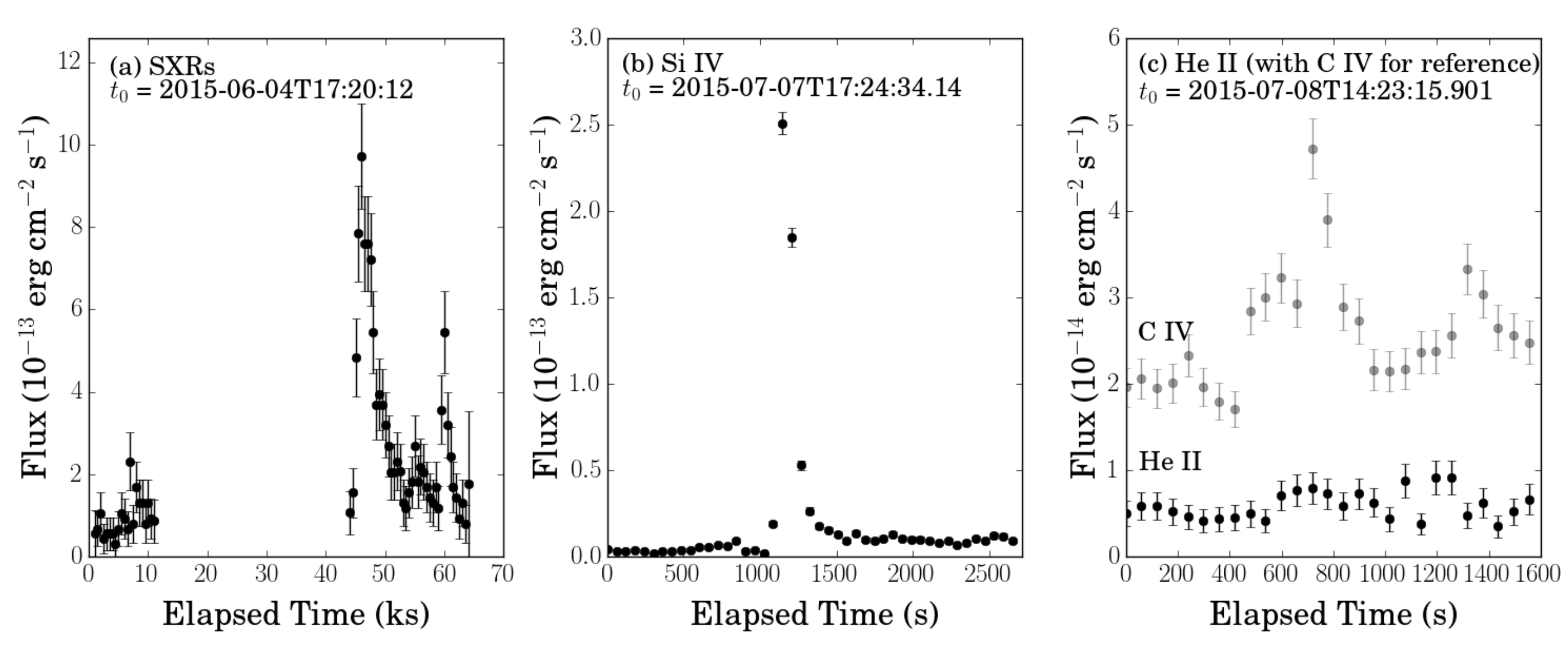} 
          }
               
   \end{center}
    \caption{A sample of observed GJ 876 flares. Panel (a) shows a \textit{Chandra}/ACIS-S3 SXR lightcurve (1.55--10.0 keV) in 500 s time bins presented originally in \cite{France2016}. Panel (b) shows a \textit{HST}/COS \SiIV~lightcurve in 60 s time bins. Panel (c) shows a \textit{HST}/COS \HeIIFUV~lightcurve (black circles) in 60 s time bins. The \CIV~lightcurve (gray circles) is shown to reference when a flare occurred.
        }
    \label{fig:GJ876_flares_fig}

\end{figure}

\subsection{Application to observed flares from GJ 876} \label{sec:app_to_flare}

The MUSCLES Treasury Survey observed several large flares from GJ 876 ($d$ = 4.7 pc) with \textit{HST} and \textit{Chandra} in 2015 June and July (\citealt{France2016,Loyd2017inprep}, in preparation; Figure~\ref{fig:GJ876_flares_fig}). Due to an \textit{HST} safing event, the \textit{Chandra} and \textit{HST} observations were not simultaneous as originally planned. On 2015 June 5, the SXR luminosity of GJ 876 was observed to increase by a factor $\textgreater$10 (Figure~\ref{fig:GJ876_flares_fig}a), and on 2015 July 7, factor of $\sim$40--100 increases were observed in the far-UV lines, including \SiIV~(Figure~\ref{fig:GJ876_flares_fig}b), and more modest increases (factor of $\sim$5) were observed in the far-UV continuum \citep{France2016}. \HeIIFUV~was observed with a different grating than \SiIV, so the two emission lines were not observed simultaneously, and no clear flare was observed in \HeIIFUV~on 2015 July 8 (Figure~\ref{fig:GJ876_flares_fig}c), although the \CIV~lightcurve indicates GJ 876 did flare during this observation. In this section and Table~\ref{table:GJ876flares}, we characterize the magnitude of these flares with the \textit{GOES} flare classification scheme and estimate the probability and magnitude of an associated particle enhancement received in GJ 876's HZ ($\langle r_{\rm HZ} \rangle$~= 0.18 AU; \citealt{Kopparapu2014}). See Section~\ref{subsubsec:limitations} for a discussion on the limitations of these results.

Table~\ref{table:GOES classification} provides comparison points for solar flares, and in Table~\ref{table:GJ876flares} we note the \textit{GOES} classification and other parameters for well-known solar and stellar flares to give context for the GJ 876 flares. The Carrington event of 1859, arguably the largest solar flare ever recorded, has been calculated to be a X45 ($\pm$5) class flare \citep{Cliver2013}. The November 4 solar flare of the Halloween storms in 2003, probably the largest flare observed during the space age, is estimated to be X30.6-class (\citealt{Kiplinger2004}; note that the \textit{GOES} detectors saturate for events between X10--X20). For the great AD Leo flare of 1985 \citep{Hawley1991}, \cite{Segura2010} estimated that a HZ planet ($\langle r_{\rm HZ} \rangle$~= 0.16 AU) would have received peak $\textgreater$10 MeV proton and SXR fluxes of 5.9$\times$10$^{8}$ pfu and 9 W m$^{-2}$, respectively. The peak SXR flux at 1 AU (0.23 W m$^{-2}$) would make this an X2300-class flare. \cite{Osten2016} found that the superflare observed from the young M dwarf binary system DG CVn was equivalent to an X600,000-class flare (60 W m$^{-2}$ at 1 AU, or 6000 W m$^{-2}$ at 0.1 AU, the approximate HZ distance). We note, however, that there is evidence that SXR--proton scaling relations should break down for such large events ($\textgreater$X10-class; \citealt{Hudson2007,Drake2013}).

We measure the peak flux for the large GJ 876 SXR flare observed at $\sim$45 ks in Figure~\ref{fig:GJ876_flares_fig}a to be (9.72$\pm$1.28)$\times$10$^{-13}$ erg cm$^{-2}$ s$^{-1}$ in the 0.3--10 keV bandpass (1.25--41 \AA). 11$\pm$1.3\% of the flare flux was emitted at energies 1.5--10 keV (1.25--8 \AA), similar to the \textit{GOES} long channel (1--8 \AA), so we find that this flare is equivalent to an M9.5-class flare (error range: M7.8--X1.1; Table~\ref{table:GOES classification}). X1-class flares have an 80\%--100\% chance of associated energetic particles from the Sun \citep{Yashiro2006}, with an estimated peak proton flux of $\sim$80 pfu at 1 AU \citep{Cliver2012a}. The SXR and proton fluxes received in GJ 876's HZ will be $\sim$30$\times$~larger: 2.8$\times$10$^{-3}$ W m$^{-2}$ and 2400 pfu, respectively. \cite{Veronig2002} find that the Sun emits X-class flares roughly every month, but flares that unleash SXR fluxes of 10$^{-3}$ W m$^{-2}$ on Earth occur only approximately once every five years. 

The smaller GJ 876 SXR flare observed at $\sim$7 ks in Figure~\ref{fig:GJ876_flares_fig}a is estimated to be equivalent to a M2.2-class flare (Table~\ref{table:GOES classification}). This smaller flare has a 40\%--80\% chance of associated energetic particles \citep{Yashiro2006} with an estimated peak proton flux $\sim$8 pfu at 1 AU. M-class flares are emitted by the Sun about every other day \citep{Veronig2002}, but flares that are a factor of 30 larger in SXR and proton fluxes, as would be experienced in GJ 876's HZ, occur only a few times a year.

GJ 876's 10$^{31}$ erg UV flare ($\Delta \lambda$~= 400--1700 \AA) observed on 2015 July 7 with \textit{HST} \citep{France2016} emitted 1.2$\times$10$^{29}$ erg in the \SiIV~emission line over $\sim$25 minutes (Figure~\ref{fig:GJ876_flares_fig}b). Using Equation~\ref{eq:prot_peak_SiIV} and the fluence at 1 AU, we find that the peak proton flux received at 1 AU during the flare was $\sim$75 pfu, or $\sim$2300 pfu at 0.18 AU. Using the SXR--UV scaling relation (Equation~\ref{eq:Xray_SiIV_peak}) and the observed \SiIV~flare peak, we find that this flare was X38-class with an estimated error of a factor of $\sim$3 (X13--X114). Note that the peak proton flux calculated for this 2015 July 7 \SiIV~flare ($\sim$2300 pfu at 0.18 AU) is similar to the peak proton flux calculated for the 2015 June 5 SXR flare ($\sim$2400 pfu at 0.18 AU), but that the \textit{GOES} classifications are different: M9.5 compared to X38, or a factor of 40 difference in SXR flux.

In total, \textit{Chandra} observed three M--X class flares in 8.25 hours, and \textit{HST} observed six comparable flares (within an order of magnitude) in 12.35 hours, including observations from the MUSCLES pilot survey \citep{France2013}. Thus, we estimate GJ 876 emits flares of this magnitude $\sim$0.4--0.5 hr$^{-1}$. From \cite{Veronig2002}, the Sun's rate of M-class flares is $\sim$0.02 hr$^{-1}$, a factor of $\sim$20 less frequent than GJ 876. However, note that these M-class flares are effectively 30$\times$ stronger (i.e. X10-class) in GJ 876's HZ at 0.18 AU. The Sun emits X10-class flares approximately once every five years, so planets in the GJ 876 HZ are receiving SXR flare fluxes $\geq$10$^{-3}$ W m$^{-2}$ associated with proton fluxes $\sim$10$^3$ pfu about four orders of magnitude more frequently than the Earth.

\begin{figure}[t]
   \begin{center}
   
     \subfigure{
          \includegraphics[width=0.6\textwidth]{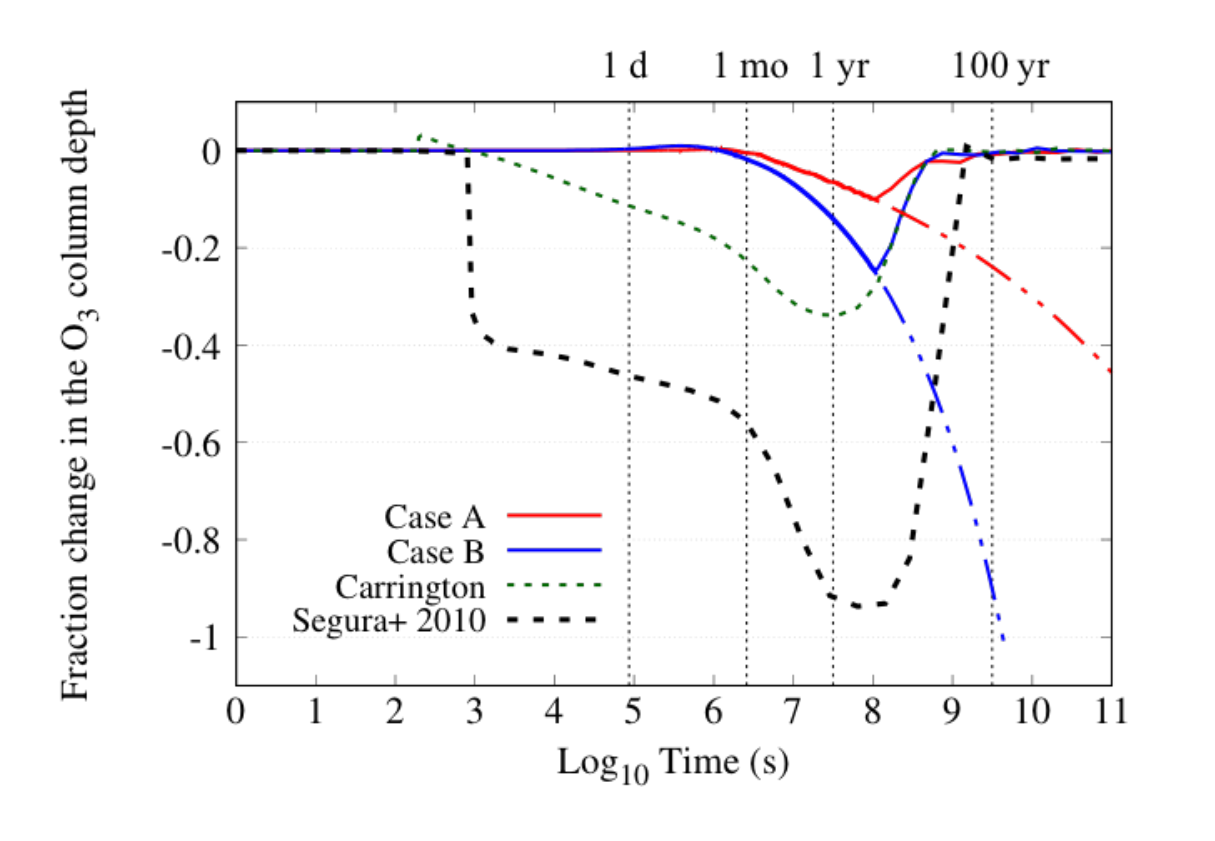}
          }
               
   \end{center}
    \caption{Fraction change in O$_3$ column depth as a function of time on a terrestrial planet with an Earth-like atmosphere and no magnetosphere (\citealt{Tilley2017inprep}, in preparation). The four vertical dotted lines denote elapsed time of 1 day, 1 month, 1 year, and 100 years, respectively. Case A (solid red line) and Case B (solid blue line) represent the O$_3$ column depth for multiple proton impact events each with $\sim$1.2$\times$10$^3$ pfu (a representative value for the flares discussed in Section~\ref{sec:app_to_flare}) for a planet orbiting GJ 876 at 0.18 AU. The frequencies of impact are 0.08 hr$^{-1}$ for Case A and 0.5 hr$^{-1}$ for Case B over a period of 40 months. The dash-dotted lines indicate a best fit to the behavior of the O$_3$ if the conditions for Case A and Case B were continued indefinitely as opposed to a period of 40 months. The dashed green line represents a single Carrington-class event (6.3$\times$10$^6$ pfu at 0.18 AU), and the dashed black line represents a single AD Leo great flare sized event (5.9$\times$10$^8$ pfu at 0.16 AU; \citealt{Rodger2008,Segura2010}). 
        }
    \label{fig:O3_proton_effect}

\end{figure}  

In Figure~\ref{fig:O3_proton_effect}, we show the effect of frequent X-class flares with associated proton enhancements on the O$_3$ column depth in an Earth-like atmosphere with no magnetosphere (\citealt{Segura2010,Tilley2017inprep}, in preparation). For comparison to the M--X class GJ 876 flares analyzed in this section, we show the dramatic responses of O$_3$ after the great AD Leo flare \citep{Segura2010} and the Carrington event (see Table~\ref{table:GJ876flares}). To represent the GJ 876 flares discussed in this section, we scaled the SED of the great AD Leo flare \citep{Hawley1991} down in intensity and duration to match flares of the dM4e star GJ 1243 \citep{Hawley2014}. This resulted in $\sim$X1-class flares with $\sim$4 minute durations. Peak proton fluxes were assigned as 1.2$\times$10$^{3}$ pfu (a typical value for the flares discussed in this section), and we provide two cases distinguished by their flare frequency: 0.08 hr$^{-1}$ (Case A) and 0.5 hr$^{-1}$ (Case B; similar to GJ 876's observed flare frequency). Both the Case A and Case B flares turn off after a period of 40 months, but via extrapolation, we show that for Case A, the O$_3$ column approaches near-complete depletion at approximately 10$^{13}$ s (318 kyr) of similar, constant stellar activity. Case B shows near-complete O$_3$ depletion after only approximately 5$\times$10$^9$ s (160 yr). Given the several Gyr period of high activity during the evolution of early-to-mid M dwarfs, both scenarios indicate massive, and likely complete, O$_3$ depletion. This suggests planetary surfaces in the HZ would be bathed in stellar UV flux. A detailed analysis of atmospheric effects will be presented in the upcoming work by \cite{Tilley2017inprep}, in preparation.

\begin{deluxetable}{lccccc}
\tablecolumns{6}
\tablewidth{0pt}
\tablecaption{GJ 876 flare properties \label{table:GJ876flares} }
\tablehead{\colhead{Flare} & 
                  \multicolumn{2}{c}{Peak SXR flux (W m$^{-2}$)} & 
                  \colhead{\textit{GOES}} &
                  \multicolumn{2}{c}{$\textgreater$10 MeV proton flux (pfu)}
                  \\
                  \colhead{} & 
                  \colhead{1 AU} & 
                  \colhead{$\langle r_{\rm HZ} \rangle$} &
                  \colhead{class} & 
                  \colhead{1 AU} & 
                  \colhead{$\langle r_{\rm HZ} \rangle$}
                  }
\startdata
GJ 876 flares: \\[4pt]
2015 June 4 & 2.2$\times$10$^{-5}$ & 6.6$\times$10$^{-4}$ & M2.2 & 8 & 240 \\[4pt]
2015 June 5 & 9.5$\times$10$^{-5}$ & 2.9$\times$10$^{-3}$ & M9.5 & 80 & 2400 \\[4pt]
2015 July 7& 3.8$\times$10$^{-3}$ & 1.1$\times$10$^{-1}$ & X38 & 75 & 2300 \\[4pt]
\hline
Reference flares: \\[4pt]
Carrington & 4.5$\times$10$^{-3}$ & -- & X45$\pm$5 & -- & -- \\
Event 1859 $^a$ & & & &\\[4pt]
2003 November 4 & 3.06$\times$10$^{-3}$ & -- & X30.6 & 400 & -- \\
solar flare $^b$ & & & & \\[4pt]
Great AD Leo & 0.23 & 9 & X2300 & 1.5$\times$10$^6$ & 5.9$\times$10$^8$ \\
flare of 1985 $^c$ & & & &  \\[4pt]
DG CVn & 60 & 6000 & X600,000 & -- &  -- \\
2014 April 24 $^d$ & & & & \\[4pt]
\enddata
\tablecomments{$\langle r_{\rm HZ} \rangle$ =  0.18 AU for GJ 876; 0.16 AU for AD Leo; 0.1 AU for DG CVn; 1 AU for the Sun.}
\tablerefs{(a) \citealt{Cliver2013}, (b) \citealt{Kiplinger2004}, (c) \citealt{Segura2010}, (d) \citealt{Osten2016}.}
\end{deluxetable}

\subsection{Limitations of the method} \label{subsubsec:limitations}

Our new method of proton flux estimation due to stellar flares has important limitations, but will be useful until advancements are made in the indirect detection of particles from stellar eruptive events (e.g., coronal dimming or Type II or III radio bursts; \citealt{Harra2016,Crosley2016}) or in our understanding of particle acceleration under kG magnetic field strengths. The scaling relations are statistical and are relatively inaccurate for individual flares.

The first caveat to the method is that we are necessarily assuming that particle acceleration in M dwarf atmospheres is similar to the Sun. This could be a poor assumption as M dwarfs have different atmospheric structure and stronger surface magnetic fields. Magnetic processes are ultimately responsible for flares and particle acceleration. Also, fast-rotating M dwarfs have extremely large surface magnetic fields with photospheres possibly saturated with active regions. Overlying magnetic fields could prevent the acceleration of particles away from the stellar atmosphere (e.g., \citealt{Vidotto2016}). This phenomenon was observed on the Sun in October 2014 when the large active region 2192 emitted many X-class flares, which have a $\textgreater$90\% probability of an associated CME, but no CMEs were ejected. Strong overlying magnetic fields were observed and have been cited as a possible explanation for the lack of associated CMEs \citep{Thalmann2015a,Sun2015}.

We should be careful when extrapolating solar-based SXR--particle scaling relations to large energies. There is evidence for a break in the SXR--proton power law around X10-class flares (see \citealt{Lingenfelter1980} and references within \citealt{Hudson2007}). It is unclear if only the expected proton flux flattens out for increasingly large SXR flares ($\textgreater$X10), or if the frequency of $\textgreater$X10-class flares also break from the expected power-law frequency distribution (e.g., \citealt{Veronig2002}). This uncertainty is partly due to the rarity of these energetic events, but also because the \textit{GOES} SXR detectors saturate around 10$^{-3}$ W m$^{-2}$. \cite{Drake2013} find that extrapolating SXR--CME scaling relations to energies applicable to highly-active solar-type stars would require CMEs to account for $\sim$10\% of the total stellar luminosity, a likely unreasonable fraction. Those authors conclude the SXR--CME scaling relations must flatten out at some point.

Predicting proton flux even for the Sun is not an easy feat, because the mechanisms of particle acceleration are not well understood. SXRs and UV photons during flares correlate with particle flux, and \cite{Kahler1982} proposes that such correlations are manifestations of ``Big Flare Syndrome", which describes how the energy of an eruptive event can power numerous physical processes that are not directly linked. Thermally heated plasma ($T$$\textgreater$80,000 K) emits short-wavelength photons that we observe as a flare, but particles are accelerated by a non-thermal process, and the ambient corona is thought to be the major contributor to CME mass \citep{Sciambi1977,Hovestadt1981}. For our purposes of estimating proton fluxes from M dwarfs and their effects on orbiting planets, a correlation between photons and particles without direct causation is valuable.

Another limitation is our use of the GJ 832 synthetic spectrum from \cite{Fontenla2016} to estimate \HeIIEUV~flare flux from far-UV \SiIV~or \HeIIFUV~flare flux. The model is applicable to the quiescent star, and flux ratios between emission lines likely change during a flare. Comparing flux ratios between the active and quiet Sun models listed in Table 1 of \cite{Fontenla2016}, the ratios of emission lines formed at different temperature change by a factor of a few during a flare. In a future work, we will improve the flux ratio relations using flare atmospheres. 

We chose \HeIIEUV~as a proxy because of its similar formation temperature to \SiIV~$\lambda \lambda$1393,1402 (and \HeIIFUV), but we note that the electrons responsible for the collisional excitation of \HeIIEUV~(40.8 eV above ground) must have significantly higher energies than the thermal electrons at $T_{\rm form}$ = 80,000 K ($k_B$$T_{\rm form}$ = 6.9 eV) that collisionally excite \SiIV~(8.8 eV above ground; \citealt{Jordan1975}). Higher energy electrons could diffuse down through the transition region as recombination/ionization timescales are much longer than dynamical timescales (e.g., \citealt{Shine1975,Golding2014,Golding2016}). Also, the \HeII~line fluxes receive some contribution from recombination, and this becomes more important during flares.

Another challenge in the analysis is quantifying the probability that any ejected particles will intersect the exoplanet. There are many unknowns here, including the interplanetary magnetic field topology that charged particle trajectories will follow \citep{Parker1958}, the opening angle of the accelerated particles, the planet's cross-section, which may be larger than $\pi R_{\rm p}^2$ due to the presence of a magnetosphere, or the direction of the particle ejection with respect to the planet's orbital plane (see \citealt{Kay2016}). A thorough treatment of this issue is beyond the scope of this work.

\section{Summary} \label{sec:Conclusions}

In this paper, we have developed methods for estimating the high-energy radiation and particle environments of M dwarfs for use when direct observations are unavailable. We have empirically determined scaling relations that can be used to estimate the UV spectra of M dwarfs from optical spectra and the energetic particle flux from UV flares. The main results of this work are summarized as follows.

\begin{enumerate}
\item Time-averaged \cak~(3933 \AA) residual equivalent width correlates positively with stellar surface flux of nine far- and near-UV emission lines, including \HI~\Lya~(Table~\ref{table:emlines}). The presented \cak~and UV scaling relations allow for the UV spectrum of any M dwarf to be approximated from ground-based optical spectra. \label{no1}
\item We present a scaling relation between \cak~equivalent width and the extreme-UV spectrum for M dwarfs based on the \Lya--extreme-UV scaling relations presented in \cite{Linsky2014} and the \cak--\Lya~scaling relation determined in this work (Table~\ref{table:CaK_EUV}). We estimate that the \ca-based extreme-UV fluxes are accurate within 40\%. \label{no2}
\item We present a new method to estimate the energetic ($\textgreater$10 MeV) proton flux emitted during far-UV emission line flares (specifically \SiIV~$\lambda \lambda$1394,1402 and \HeII~$\lambda$1640; Table~\ref{table:prot_correlations}). The UV--proton scaling relations are derived from solar irradiance observations and \textit{in situ} proton measurements near Earth. \label{no3}
\item We present methods to estimate the \textit{GOES} flare classification corresponding to far-UV \SiIV~and \HeII~flares (Equations~\ref{eq:Xray_SiIV_peak} and \ref{eq:Xray_HeIIFUV_peak}). This is important for estimating the probability that any far-UV flare will have an accompanying proton enhancement. Larger solar flares have a higher probability of associated particle enhancements. \label{no4}
\item We analyze several flares observed with \textit{Chandra} and \textit{Hubble} from the M4 dwarf GJ 876 as part of the MUSCLES Treasury Survey, and place the flare properties in context with solar flares (Section~\ref{sec:app_to_flare}). We find that planets in GJ 876's HZ experience large, Carrington-like flares (soft X-ray flux $\textgreater$10$^{-3}$ W m$^{-2}$) and particle enhancements approximately four orders of magnitude more frequently than Earth. Flare activity at the level observed on GJ 876 is predicted to lead to complete stripping of O$_3$ from an Earth-like planet on timescales between 10$^2$ and 10$^5$ years. \label{no5}
\end{enumerate}

\noindent M dwarfs are currently a prime target for exoplanet searches and characterization efforts with current and near-future technologies including \textit{JWST}, but important questions have been raised about these planets' potential habitability. \textit{JWST} will be able to characterize only a few M dwarf exoplanets, and we need to ensure that HZ terrestrial planets orbiting nearby M dwarfs with the maximum probability of hosting a detectable atmosphere are chosen for \textit{JWST} target selection. The stellar host's UV SED and energetic particle flux are important drivers of chemistry, heating, and mass loss, but we do not have the space telescope resources for reconnaissance of every nearby M dwarf system, and energetic particles cannot be measured directly across interstellar distances. In this paper, we have provided tools to estimate UV SEDs from comparatively easy-to-obtain \ca~measurements and energetic particle fluxes from observed UV and X-ray flares.

\acknowledgements
The data presented here were obtained as part of the \textit{HST} Guest Observing programs \#12464 and \#13650 as well as the COS Science Team Guaranteed Time programs \#12034 and \#12035. This work was supported by NASA grants HST-GO-12464.01 and HST-GO- 13650.01 to the University of Colorado at Boulder. This work is also based on data from the Chandra X-ray Observatory (ObsIDs: 17315 and 17316) and supported by CXO grant G05-16155X. Observations were also obtained with the Apache Point Observatory 3.5-meter telescope, which is owned and operated by the Astrophysical Research Consortium. This research has made extensive use of the Keck Observatory Archive (KOA), which is operated by the W. M. Keck Observatory and the NASA Exoplanet Science Institute (NExScI), under contract with the National Aeronautics and Space Administration. The ESO Science Archive Facility was also used to access VLT/XSHOOTER science products. This paper includes data gathered with the 6.5-m Magellan Telescopes located at Las Campanas Observatory, Chile. AY thanks Don Woodraska for assistance with the \textit{GOES} and \textit{SDO}/EVE data, Juan Fontenla for providing the GJ 832 synthetic spectrum, and Lauren W. Blum for helpful discussions. ERN acknowledges support from an NSF Astronomy and Astrophysics Postdoctoral Fellowship under award AST-1602597, and AS acknowledges the support of UNAM-PAPIIT project IN109015.

\facilities{HST (COS, STIS), ARC (DIS), CXO, Keck:I (HIRES), CASLEO:JST (REOSC), VLT:Kueyen (XSHOOTER), Magellan:Clay (MIKE), SDO (EVE), GOES.}
\software{IRAF, Astropy \citep{Robitaille2013}, IPython \citep{Perez2007}, Matplotlib \citep{Hunter2007}, Pandas \citep{McKinney2010}, NumPy and SciPy \citep{VanderWalt2011}.}

\bibliography{../../Mendeley}{}
\bibliographystyle{aasjournal}

\appendix
\section{Equivalent Width Correction} \label{app: Though Experiment}

\begin{figure}[t]
   \begin{center}
   
     \subfigure{
          \includegraphics[width=\textwidth]{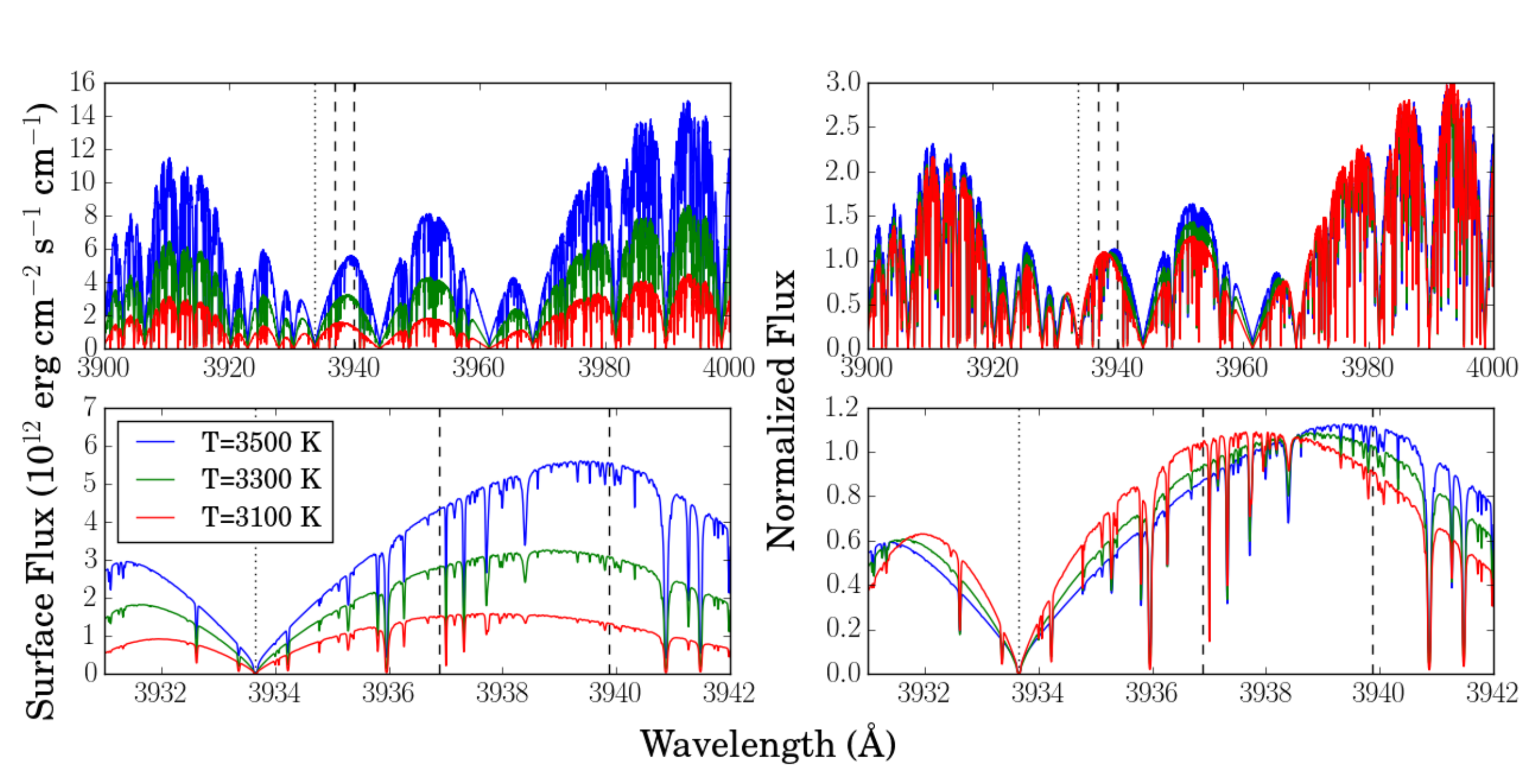}
          }
          
   \end{center}
    \caption{\cak~photospheric absorption for 3 different M dwarf PHOENIX models -- 3100 K, 3300 K, and 3500 K \citep{Husser2013}. All models have log$_{10}$ $g$ = 5 and [Fe/H] = 0 (solar). The vertical dotted black line shows the \cak~line center (3933.66 \AA), and the two vertical dashed lines show the continuum region (3936.9--3939.9 \AA). The two left panels show the surface flux for the three different models, and the two right panels show the same models normalized to the average flux value in the continuum region.
        }
    \label{fig:phx_spectra_appendix}

\end{figure}

The following exemplifies how a spectral type dependence is introduced in \cak~equivalent width measurements, and how it can be removed. See Figure~\ref{fig:phx_spectra_appendix} for examples on how the PHOENIX models used in Section~\ref{sec:eqw} change as a function of spectral type.

Two stars, A and B, have different spectral types ($T_{\rm A}$ $\textgreater$~$T_{\rm B}$ and $R_{\rm A}$ $\textgreater$~$R_{\rm B}$) and are observed at different distances, $d_{\rm A}$ and $d_{\rm B}$, from Earth. These stars have the same \ca~emission core surface flux (erg cm$^{-2}$ s$^{-1}$), $F_{\rm A, S}$ = $F_{\rm B, S}$, and therefore should have the same \cak~equivalent widths, $W_{\rm \lambda}$. From flux-calibrated \ca~spectra, we perfectly isolate the emission core from the observed line profile by fitting and subtracting a radiative equilibrium model, and we integrate over the residual \cak~emission, finding $F_{\rm A, S}$ $\times$ $\frac{R_{\rm A}^2}{d_{\rm A}^2}$ and $F_{\rm B, S}$ $\times$ $\frac{R_{\rm B}^2}{d_{\rm B}^2}$ for Star A and Star B, respectively. 

To convert the observed fluxes into equivalent widths, we normalize to nearby continuum, which has an average surface flux $x$($T$). Because Star A is warmer, $x$($T_{\rm A}$)~$\textgreater$~$x$($T_{\rm B}$), and the resulting equivalent widths are not equal:

\begin{equation}
W_{\rm \lambda, A} = \frac{ F_{\rm A, S} \times~\frac{R_{\rm A}^2}{d_{\rm A}^2}}{x(T_{\rm A}) \times~\frac{R_{\rm A}^2}{d_{\rm A}^2}} = \frac{ F_{\rm A, S}}{x(T_{\rm A})},
\label{eq:eqwA}
\end{equation}

\noindent and

\begin{equation}
W_{\rm \lambda, B} = \frac{ F_{\rm B, S} \times~\frac{R_{\rm B}^2}{d_{\rm B}^2}}{x(T_{\rm B}) \times~\frac{R_{\rm B}^2}{d_{\rm B}^2}} = \frac{ F_{\rm B, S}}{x(T_{\rm B})}.
\label{eq:eqwB}
\end{equation}

To correct the equivalent widths so that they are equal for Star A and Star B, we multiply by $x$($T$) scaled to a reference value $x$($T_{\rm ref}$). The ``corrected" equivalent width is given by: 

\begin{equation}
W_{\rm \lambda, corr} = W_{\rm \lambda} \times~\frac{x(T)}{x(T_{\rm ref})}.
\label{eq:eqwcorr}
\end{equation}

Figure~\ref{fig:phx_spectra_appendix} shows radiative equilibrium model spectra (PHOENIX; \citealt{Husser2013}) for 3 spectral subtypes from our sample -- 3100 K, 3300 K, and 3500 K. The cooler spectral subtypes have narrower \ca~absorption, as well as fainter continuum.

\section{UV--UV correlations} \label{app:uvuv}

Here we present scaling relations between the nine far- and near-UV emission lines presented in Tables~\ref{table:UVfluxes1}, \ref{table:UVfluxes2}, and \ref{table:emlines}, and between these UV emission lines and the stellar rotation periods presented in Table~\ref{table:targets}. See Section~\ref{sec:UVopticaldata} for a discussion of the UV spectra, including \Lya~reconstructions and other corrections made for ISM absorption. All observed fluxes have been converted to surface fluxes using the stellar radii and distances presented in Table~\ref{table:targets}.

In Figures~\ref{fig:UV-UV1}--\ref{fig:UV-UV4}, we show the positive correlations between UV surface fluxes ($F_{\rm S, UV1}$, $F_{\rm S, UV2}$) with their power law fits of the form log$_{10}$ $F_{\rm S, UV1}$ = m $\times$~log$_{10}$ $F_{\rm S, UV2}$ + b. We find statistically significant correlations for all 36 combinations of the nine UV emission lines. The Pearson correlation coefficients range from 0.80--0.99, and the scatter about the best-fit lines ranges from 0.11--0.53 dex.

\begin{figure}
   \begin{center}
   
     \subfigure{
          \includegraphics[width=\textwidth]{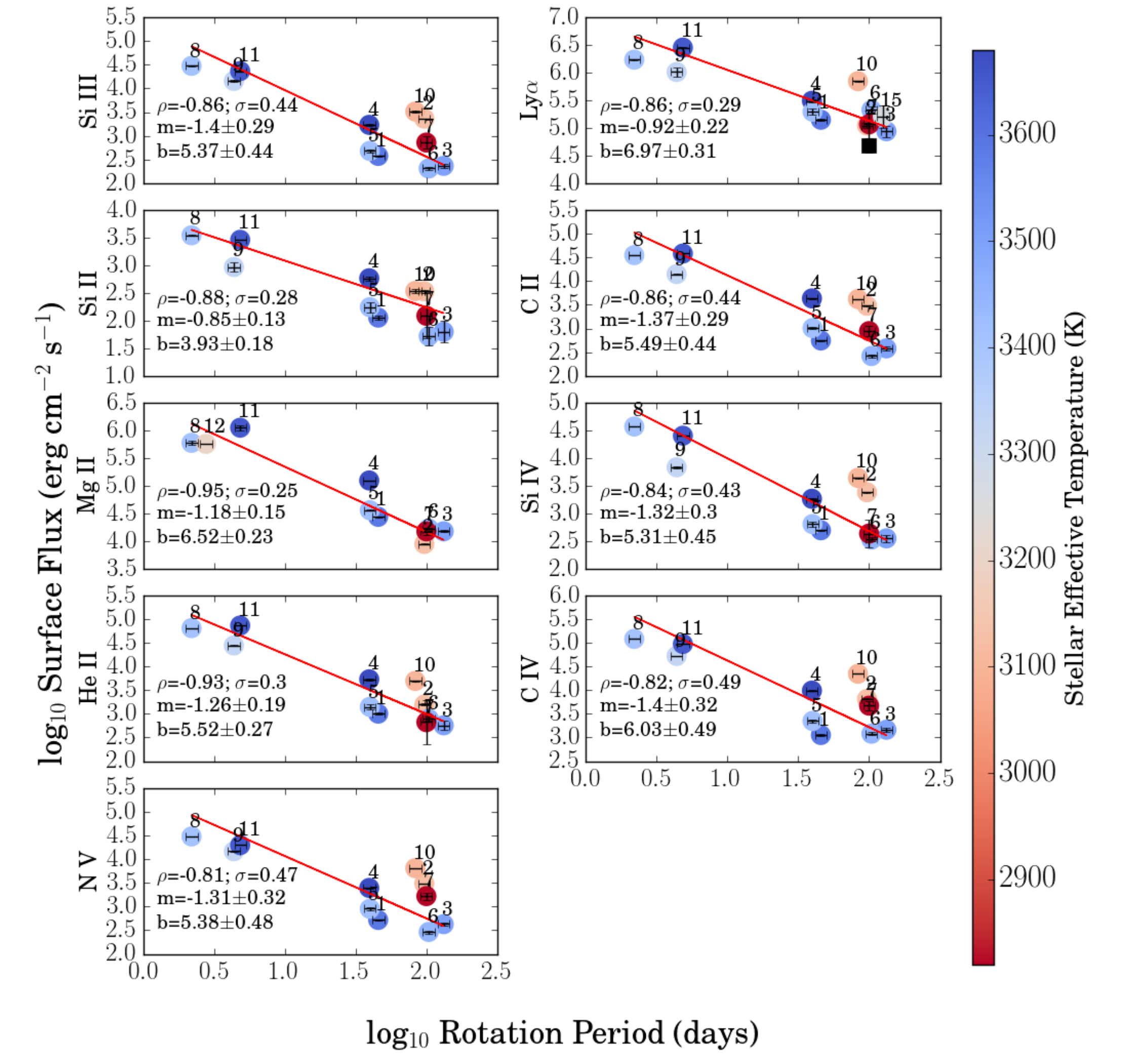}
          }
          
   \end{center}
    \caption{Surface fluxes (erg cm$^{-2}$ s$^{-1}$) of the UV emission lines (see Tables~\ref{table:UVfluxes1}, \ref{table:UVfluxes2}, and \ref{table:emlines}) and the stellar rotation periods (days; Table~\ref{table:targets}). 10\% error bars have been placed on the rotation periods. Each point represents a different star (numbered to match Table~\ref{table:targets}) color-coded by effective temperature. The black square in the top right panel represents the uncorrected \Lya~surface flux for GJ 1214 (see \citealt{Youngblood2016}); GJ 1214 was not included in that panel's fit. The data were fitted by power laws of the form log$_{10}$ $F_{\rm S,UV}$ = m $\times$~log$_{10}$ $P_{\rm rot}$ + b, where $F_{\rm S,UV}$ is the surface flux of the UV emission line in erg cm$^{-2}$ s$^{-1}$, and $P_{\rm rot}$ is the rotation period in days. The fit coefficients m and b, the Pearson correlation coefficient $\rho$, and the standard deviation of the data points about the best-fit line $\sigma$ (dex) are printed in each panel.
        }
    \label{fig:UV-UV5}

\end{figure}

In Figure~\ref{fig:UV-UV5}, we present the negative correlations between stellar rotation period ($P_{\rm rot}$) and the nine UV emission lines with their power law fits of the form log$_{10}$ $F_{\rm S, UV1}$ = m $\times$~log$_{10}$ $P_{\rm rot}$ + b. All of the correlations are statistically significant, although our M dwarf sample is divided between two widely-separated populations: $P_{\rm rot}$~$\textless$~5 days and $P_{\rm rot}$~$\textgreater$~39 days. Within the slowly-rotating population or the fast-rotating population, most of the correlations are not statistically significant.

\cite{Youngblood2016} presented scaling relations between \Lya~and \MgII, \SiIII, \CIV, and $P_{\rm rot}$ for a smaller sample of M dwarfs. The shallow correlations (m $\textless$~0.15) found between \Lya~and \SiIII, \CIV~were not statistically significant and were consistent with zero, likely due to the small sample size and the small range of explored parameter space. This work's power law exponents between \Lya~and \SiIII, \CIV~are m = 0.60$\pm$0.09 and m = 0.63$\pm$0.08, respectively, compared to the m = 0.07$\pm$0.31 and m = 0.13$\pm$0.35 values found by \cite{Youngblood2016}. For \Lya~and \MgII, our power law exponent m = 0.73$\pm$0.08 is consistent with the m = 0.77$\pm$0.10 value from \cite{Youngblood2016}, as is our exponent for the \Lya~and $P_{\rm rot}$ relation: m = -0.92$\pm$0.22 compared to m = -0.86$\pm$0.16.

\begin{figure}
   \begin{center}
   
     \subfigure{
          \includegraphics[width=\textwidth]{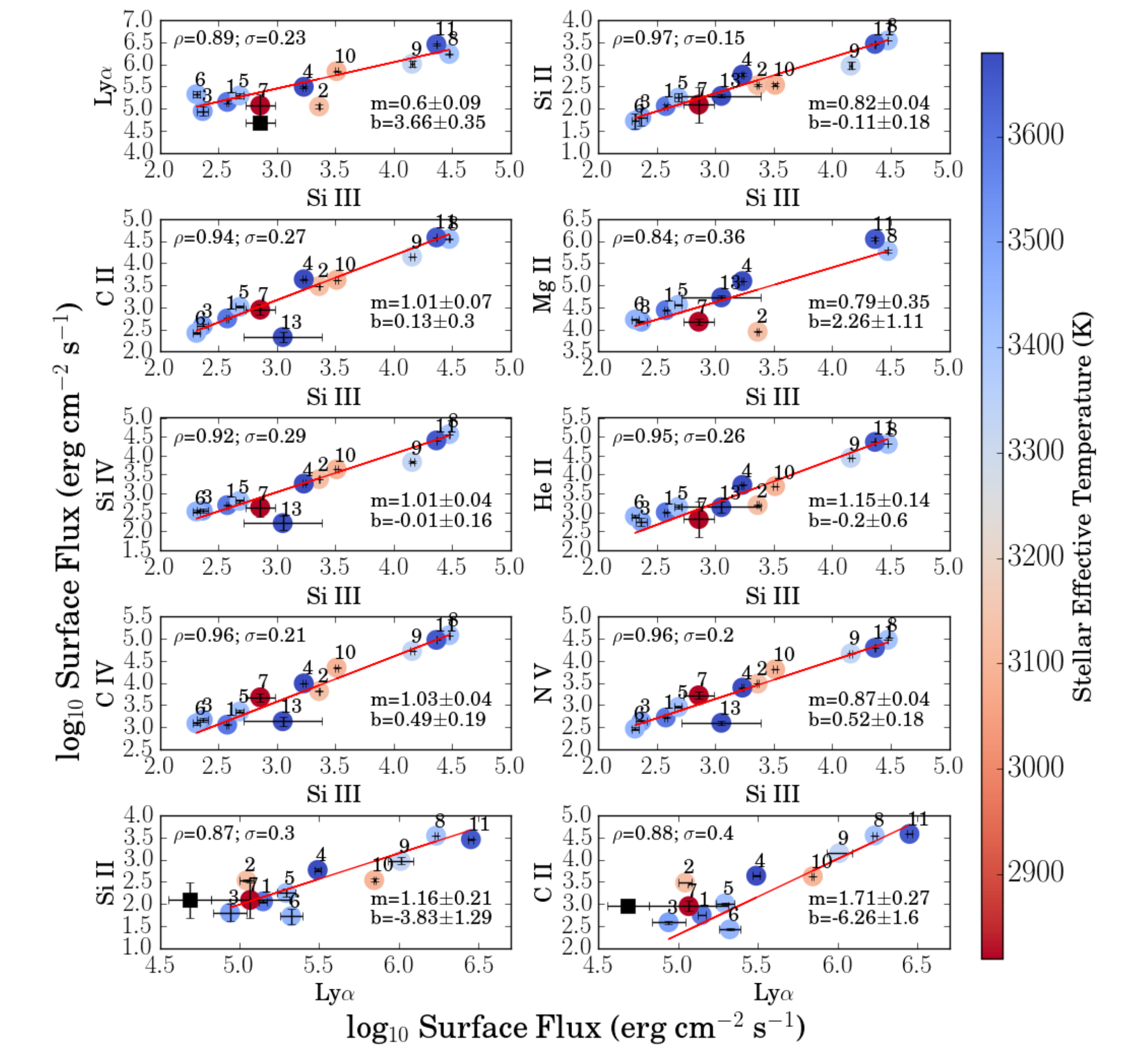}
          }
          
   \end{center}
    \caption{Surface fluxes (erg cm$^{-2}$ s$^{-1}$) for the individual UV emission lines (see Tables~\ref{table:UVfluxes1}, \ref{table:UVfluxes2}, and \ref{table:emlines}). Each point represents a different star (numbered to match Table~\ref{table:targets}) color-coded by effective temperature. The black squares in the panels showing \Lya~flux represent the uncorrected \Lya~surface flux for GJ 1214 (see \citealt{Youngblood2016}). The data were fitted by power laws of the form log$_{10}$ $F_{\rm S,UV1}$ = m $\times$~log$_{10}$ $F_{\rm S,UV1}$ + b, where $F_{\rm S,UV}$ is the surface flux of the UV emission line in erg cm$^{-2}$ s$^{-1}$. The fit coefficients m and b, the Pearson correlation coefficient $\rho$, and the standard deviation of the data points about the best-fit line $\sigma$ (dex) are printed in each panel. GJ 1214 (star 7) was not included in any of the \Lya~fits, but was included in all others.
        }
    \label{fig:UV-UV1}

\end{figure}  

\begin{figure}
   \begin{center}
   
     \subfigure{
          \includegraphics[width=\textwidth]{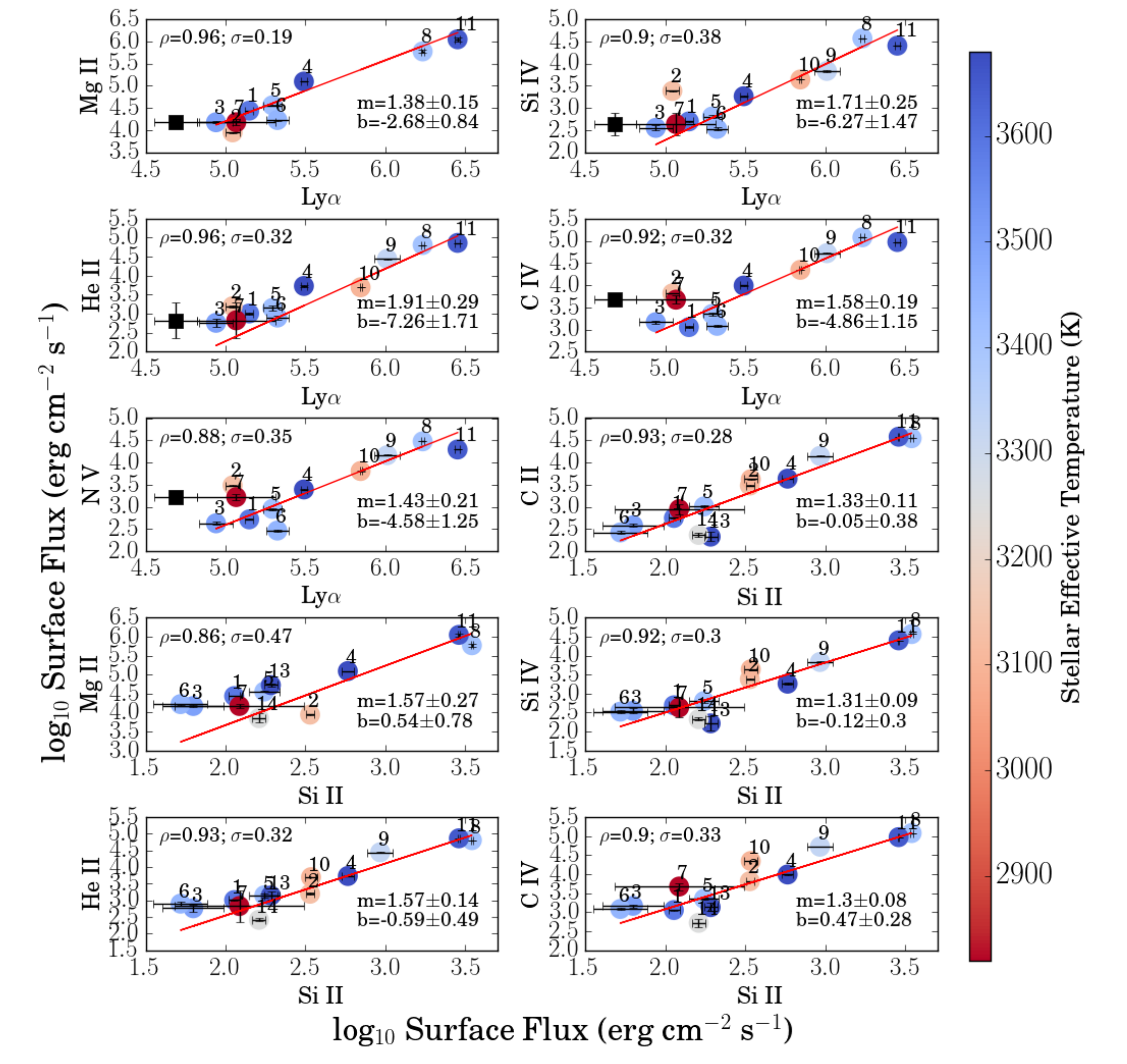}
          }
          
   \end{center}
    \caption{Continued from Figure~\ref{fig:UV-UV1}.
        }
    \label{fig:UV-UV2}

\end{figure}  

\begin{figure}
   \begin{center}
   
     \subfigure{
          \includegraphics[width=\textwidth]{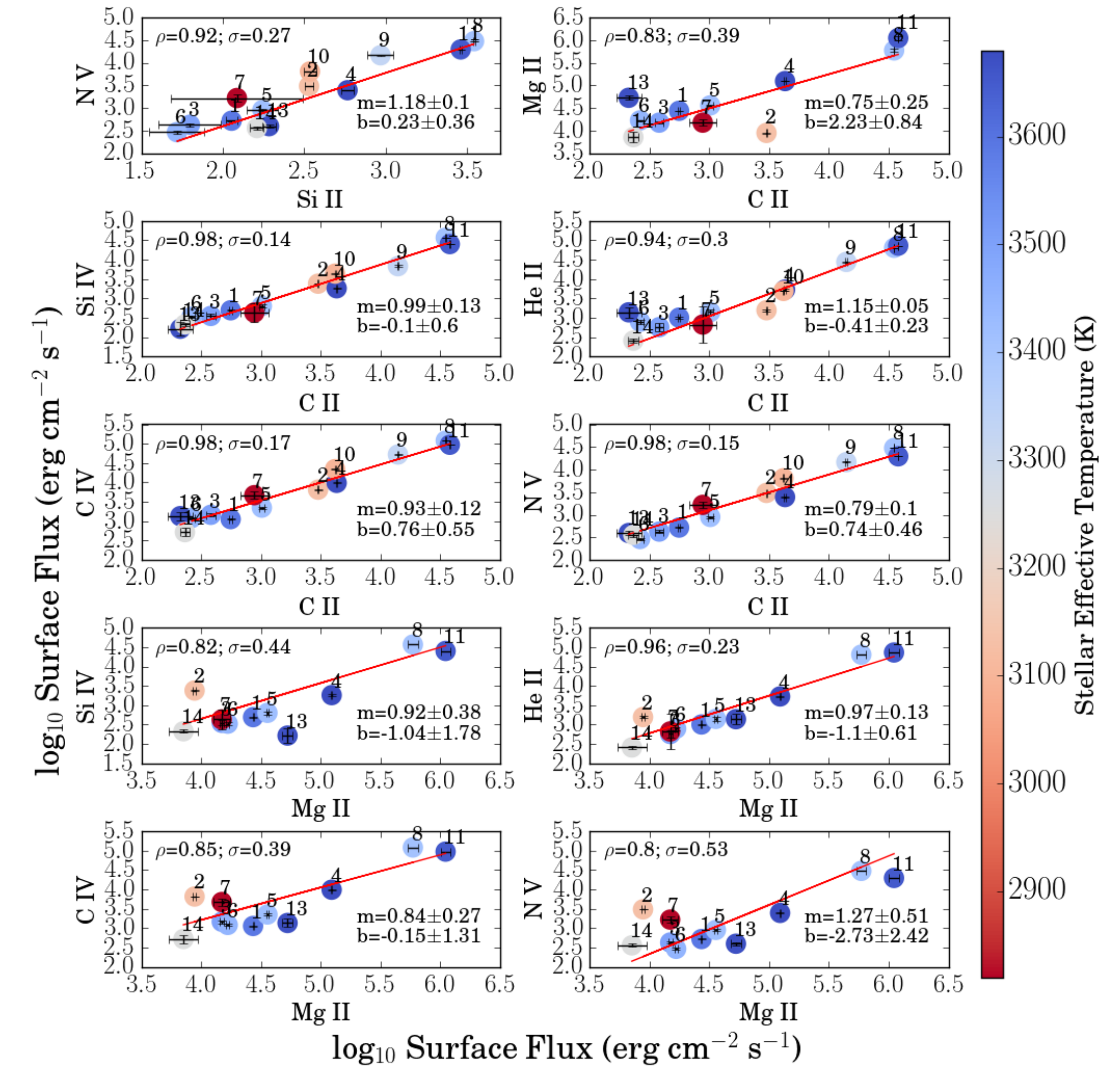}
          }
          
   \end{center}
    \caption{Continued from Figure~\ref{fig:UV-UV1}.
        }
    \label{fig:UV-UV3}

\end{figure}  

\begin{figure}
   \begin{center}
   
     \subfigure{
          \includegraphics[width=\textwidth]{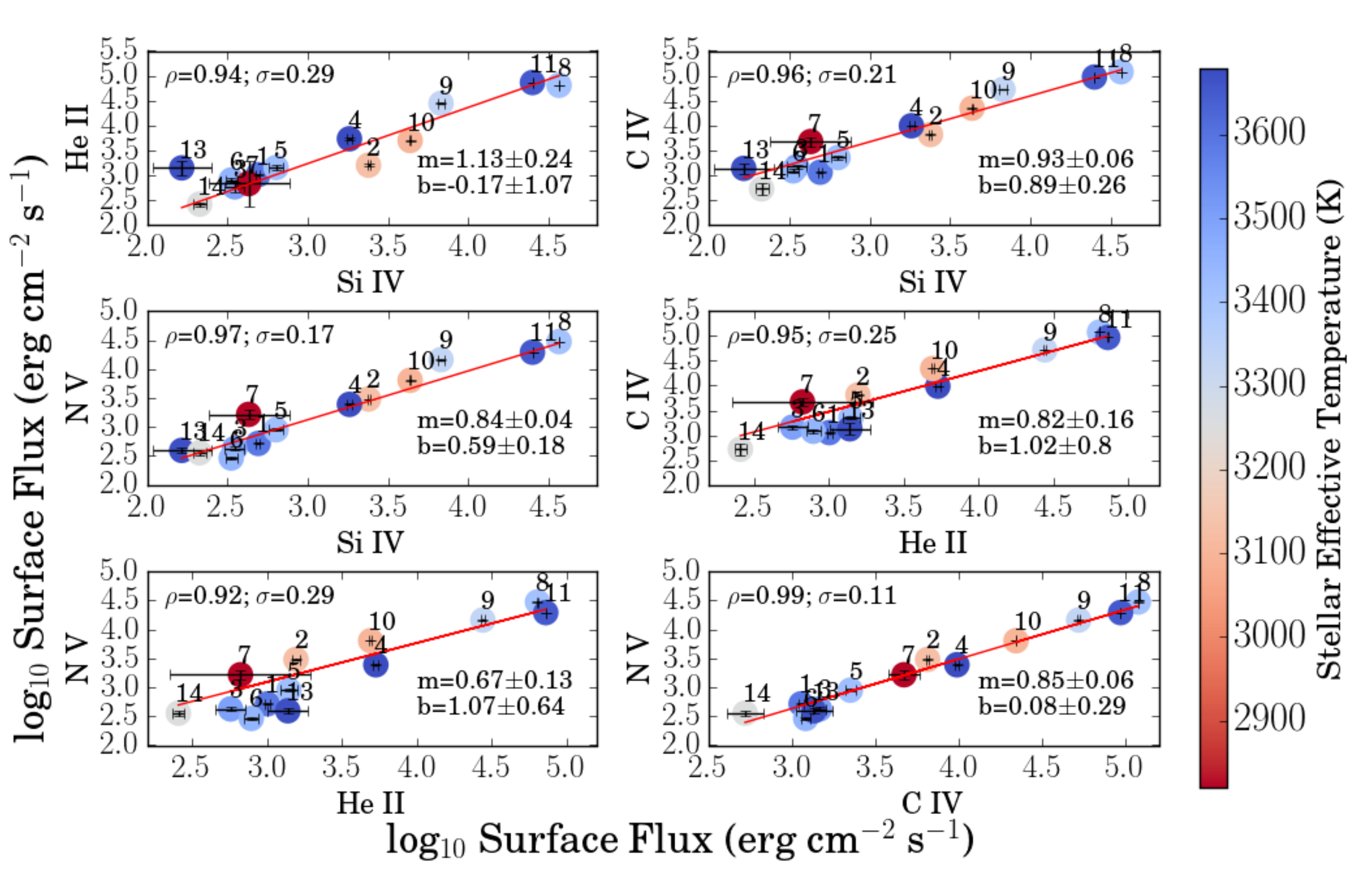}
          }
          
   \end{center}
    \caption{Continued from Figure~\ref{fig:UV-UV1}.
        }
    \label{fig:UV-UV4}

\end{figure}  

\begin{deluxetable}{cccccccc}
\tablecolumns{8}
\tablewidth{0pt}
\tablecaption{UV--UV emission line scaling relations \label{table:uv-uv} }
\tablehead{\colhead{UV1, UV2} & 
                  \colhead{m} &
                  \colhead{$\sigma_{\rm m}$} &
                  \colhead{b} & 
                  \colhead{$\sigma_{\rm b}$} & 
                  \colhead{$\rho$} & 
                  \colhead{$n$} &
                  \colhead{$\sigma$}
                  }
\startdata
Ly$\alpha$, Si III & 0.60 & 0.09 & 3.66 & 0.35 & 0.89 & 6.4E-04 & 0.23 \\
Si II, Si III & 0.82 & 0.04 & -0.11 & 0.18 & 0.97 & 2.3E-07 & 0.15 \\
C II, Si III & 1.01 & 0.07 & 0.13 & 0.30 & 0.94 & 8.1E-06 & 0.27 \\
Mg II, Si III & 0.79 & 0.35 & 2.26 & 1.11 & 0.84 & 2.2E-03 & 0.36 \\
Si IV, Si III & 1.01 & 0.04 & -0.01 & 0.16 & 0.92 & 1.7E-05 & 0.29 \\
He II, Si III & 1.15 & 0.14 & -0.2 & 0.6 & 0.95 & 1.6E-06 & 0.26 \\
C IV, Si III & 1.03 & 0.04 & 0.49 & 0.19 & 0.96 & 7.5E-07 & 0.21 \\
N V, Si III & 0.87 & 0.04 & 0.52 & 0.18 & 0.96 & 1.0E-06 & 0.20 \\
Si II, Ly$\alpha$ & 1.16 & 0.21 & -3.83 & 1.29 & 0.87 & 1.1E-03 & 0.30 \\
C II, Ly$\alpha$ & 1.71 & 0.27 & -6.26 & 1.60 & 0.88 & 7.4E-04 & 0.40 \\
Mg II, Ly$\alpha$ & 1.38 & 0.15 & -2.68 & 0.84 & 0.96 & 1.1E-04 & 0.19 \\
Si IV, Ly$\alpha$ & 1.71 & 0.25 & -6.27 & 1.47 & 0.90 & 4.1E-04 & 0.38 \\
He II, Ly$\alpha$ & 1.91 & 0.29 & -7.26 & 1.71 & 0.96 & 1.6E-05 & 0.32 \\
C IV, Ly$\alpha$ & 1.58 & 0.19 & -4.86 & 1.15 & 0.92 & 2.0E-04 & 0.32 \\
N V, Ly$\alpha$ & 1.43 & 0.21 & -4.58 & 1.25 & 0.88 & 9.1E-04 & 0.35 \\
C II, Si II & 1.33 & 0.11 & -0.05 & 0.38 & 0.93 & 4.1E-06 & 0.28 \\
Mg II, Si II & 1.57 & 0.27 & 0.54 & 0.78 & 0.86 & 6.0E-04 & 0.47 \\
Si IV, Si II & 1.31 & 0.09 & -0.12 & 0.30 & 0.92 & 1.1E-05 & 0.30 \\
He II, Si II & 1.57 & 0.14 & -0.59 & 0.49 & 0.93 & 3.7E-06 & 0.32 \\
C IV, Si II & 1.3 & 0.08 & 0.47 & 0.28 & 0.9 & 2.4E-05 & 0.33 \\
N V, Si II & 1.18 & 0.10 & 0.23 & 0.36 & 0.92 & 1.0E-05 & 0.27 \\
Mg II, C II & 0.75 & 0.25 & 2.23 & 0.84 & 0.83 & 1.7E-03 & 0.39 \\
Si IV, C II & 0.99 & 0.13 & -0.10 & 0.60 & 0.98 & 1.6E-09 & 0.14 \\
He II, C II & 1.15 & 0.05 & -0.41 & 0.23 & 0.94 & 1.3E-06 & 0.30 \\
C IV, C II & 0.93 & 0.12 & 0.76 & 0.55 & 0.98 & 1.4E-08 & 0.17 \\
N V, C II & 0.79 & 0.10 & 0.74 & 0.46 & 0.98 & 3.5E-09 & 0.15 \\
Si IV, Mg II & 0.92 & 0.38 & -1.04 & 1.78 & 0.82 & 2.1E-03 & 0.44 \\
He II, Mg II & 0.97 & 0.13 & -1.10 & 0.61 & 0.96 & 4.1E-06 & 0.23 \\
C IV, Mg II & 0.84 & 0.27 & -0.15 & 1.31 & 0.85 & 8.4E-04 & 0.39 \\
N V, Mg II & 1.27 & 0.51 & -2.73 & 2.42 & 0.80 & 3.4E-03 & 0.53 \\
He II, Si IV & 1.13 & 0.24 & -0.17 & 1.07 & 0.94 & 1.9E-06 & 0.29 \\
C IV, Si IV & 0.93 & 0.06 & 0.89 & 0.26 & 0.96 & 1.1E-07 & 0.21 \\
N V, Si IV & 0.84 & 0.04 & 0.59 & 0.18 & 0.97 & 5.9E-08 & 0.17 \\
C IV, He II & 0.82 & 0.16 & 1.02 & 0.80 & 0.95 & 6.3E-07 & 0.25 \\
N V, He II & 0.67 & 0.13 & 1.07 & 0.64 & 0.92 & 7.4E-06 & 0.29 \\
N V, C IV & 0.85 & 0.06 & 0.08 & 0.29 & 0.99 & 4.9E-10 & 0.11 \\
\enddata
\tablecomments{Scaling relations are power laws of the form log$_{10}$ $F_{\rm S, UV1}$ = m $\times$~log$_{10}$ $F_{\rm S, UV2}$ + b, where $F_{\rm S}$ is the surface flux in erg cm$^{-2}$ s$^{-1}$, $\rho$~is the Pearson correlation coefficient, $n$ is the probability of no correlation, and $\sigma$~is the standard deviation of the data points about the best-fit line (dex).}
\end{deluxetable}

\begin{deluxetable}{cccccccc}

\tablecolumns{8}
\tablewidth{0pt}
\tablecaption{UV emission line scaling relations with stellar rotation period \label{table:rot-uv} }
\tablehead{\colhead{UV} & 
                  \colhead{m} &
                  \colhead{$\sigma_{\rm m}$} &
                  \colhead{b} & 
                  \colhead{$\sigma_{\rm b}$} & 
                  \colhead{$\rho$} & 
                  \colhead{$n$} &
                  \colhead{$\sigma$}
                  }
\startdata
Si III & -1.40 & 0.29 & 5.37 & 0.44 & -0.86 & 7.3E-04 & 0.44 \\
Ly$\alpha$ & -0.92 & 0.22 & 6.97 & 0.31 & -0.86 & 7.5E-04 & 0.29 \\
Si II & -0.85 & 0.13 & 3.93 & 0.18 & -0.88 & 3.8E-04 & 0.28 \\
C II & -1.37 & 0.29 & 5.49 & 0.44 & -0.86 & 8.0E-04 & 0.44 \\
Mg II & -1.18 & 0.15 & 6.52 & 0.23 & -0.95 & 3.0E-05 & 0.25 \\
Si IV & -1.32 & 0.30 & 5.31 & 0.45 & -0.84 & 1.1E-03 & 0.43 \\
He II & -1.26 & 0.19 & 5.52 & 0.27 & -0.93 & 4.5E-05 & 0.30 \\
C IV & -1.40 & 0.32 & 6.03 & 0.49 & -0.82 & 1.9E-03 & 0.49 \\
N V & -1.31 & 0.32 & 5.38 & 0.48 & -0.81 & 2.5E-03 & 0.47 \\
\enddata
\tablecomments{Scaling relations are power laws of the form log$_{10}$ $F_{\rm S, UV}$ = m $\times$~log$_{10}$ $P_{\rm rot}$ + b, where $F_{\rm S}$ is the surface flux in erg cm$^{-2}$ s$^{-1}$ and $P_{\rm rot}$ is the rotation period in days, $\rho$~is the Pearson correlation coefficient, $n$ is the probability of no correlation, and $\sigma$~is the standard deviation of the data points about the best-fit line (dex).}
\end{deluxetable}

\end{document}